\documentclass[prl,twocolumn,nopacs,amsmath,amssymb,superscriptaddress]{revtex4-2}

\usepackage{amsmath}
\usepackage{amssymb}
\usepackage{amsbsy}
\usepackage{graphicx}
\usepackage{xcolor}

\usepackage{mathrsfs}

\usepackage{enumerate}
\usepackage{mathtools}

\usepackage{dcolumn}

\newcommand{\moy}[1]{\left\langle #1 \right\rangle}
\newcommand{\dd}[0]{\mathrm{d}}

\newcommand{\erf}[0]{\text{erf}}

\usepackage{soul}

\newcommand{\dt}[2]{\ensuremath{\frac{\dd #1}{\dd #2}}}

\def\e{e}
\def\I{\mathrm{i}}

\newcommand{\abs}[1]{\ensuremath{\left| #1 \right|}}

\DeclareMathOperator{\erfc}{erfc}

\definecolor{darkblue}{rgb}{0,0,0.6}
\definecolor{darkred}{rgb}{0.6,0,0}

\usepackage[colorlinks=true,urlcolor=darkblue,citecolor=darkblue,linkcolor=darkred]{hyperref}

\def\qt{\tilde{q}}
\def\pt{\tilde{p}}

\def\rb{\bar\rho}

\def\D{\mathcal{D}}

\def\fm{f_{\mathrm{M}}}

\def\rhom{\rho_0}

\def\Nsc{\Lambda}

\begin{document}

\title{Exact large-scale correlations in diffusive systems with general interactions:\texorpdfstring{\\}{} explicit characterisation without the Dean--Kawasaki equation}

\author{Aur\'elien Grabsch}
\affiliation{Sorbonne Universit\'e, CNRS, Laboratoire de Physique Th\'eorique de la Mati\`ere Condens\'ee (LPTMC), 4 Place Jussieu, 75005 Paris, France}

\author{Davide Venturelli}
\affiliation{Sorbonne Universit\'e, CNRS, Laboratoire de Physique Th\'eorique de la Mati\`ere Condens\'ee (LPTMC), 4 Place Jussieu, 75005 Paris, France}
\affiliation{Sorbonne Universit\'e, CNRS, Laboratoire Jean Perrin (LJP), 4 Place Jussieu, 75005 Paris, France}

\author{Olivier B\'enichou}
\affiliation{Sorbonne Universit\'e, CNRS, Laboratoire de Physique Th\'eorique de la Mati\`ere Condens\'ee (LPTMC), 4 Place Jussieu, 75005 Paris, France}

\begin{abstract}
Characterising the statistical properties of classical interacting particle systems is a long-standing question. For Brownian particles the microscopic density obeys a stochastic evolution equation, known as the Dean--Kawasaki equation. This equation remains mostly formal and linearization (or higher-order expansions) is required to obtain explicit expressions for physical observables, with a range of validity not easily defined. Here, by combining macroscopic fluctuation theory with equilibrium statistical mechanics, we provide a systematic alternative to the Dean--Kawasaki framework to characterize large-scale correlations.
This approach enables us to obtain explicit and exact results for the large-scale behavior of dynamical observables such as tracer cumulants and bath-tracer correlations in one dimension, both in and out of equilibrium.
In particular, we reveal a generic non-monotonic spatial structure in the response of the bath following a temperature quench. Our approach applies to a broad class of interaction potentials and extends naturally to higher dimensions.
\end{abstract}

\maketitle

\let\oldaddcontentsline\addcontentsline
\renewcommand{\addcontentsline}[3]{}

\emph{Introduction.---} Characterizing correlations in classical systems of interacting particles, both at equilibrium and out of equilibrium, remains a central challenge in statistical physics, with implications for systems as diverse as colloidal suspensions, active matter, and biological fluids~\cite{Hansen:2005,Doi:1988,Stribeck:2007}.
Correlations can be approximated through closure schemes in both static and dynamical regimes~\cite{Hansen:2005,Gotze:2009,Pihlajamaa:2024}.
However, a key obstacle is the lack of analytical frameworks capable of bridging microscopic dynamics and large-scale behavior without such uncontrolled approximations.

The paradigmatic model of pairwise interacting Brownian particles, which can represent various systems such as colloidal suspensions, supercooled liquids, polymers, or ions in solution, illustrates this challenge. The positions $x_i$ of the particles obey a set of overdamped Langevin equations, written here in 1D for simplicity,
\begin{equation}
    \label{eq:EqBrownianPart}
    \dt{x_i}{t} = - \mu_0 \sum_{j \neq i} V'(x_i-x_j)
    + \sqrt{2 D_0} \: \eta_i
    \:,
\end{equation}
where $\mu_0$ is the bare mobility of a particle, $V$ the interaction potential, $\eta_i$ a Gaussian white noise with unit variance, and $D_0 = \mu_0 k_B T$ the bare diffusion coefficient.

The microscopic density field $\rhom(x,t) = \sum_i \delta(x - x_i(t))$, which is the basic quantity to characterize various observables, either at equilibrium (static density correlations) or out of equilibrium (dynamical correlations, tracer displacement, currents, ...), follows a stochastic partial differential equation, known as the Dean--Kawasaki equation (DKE)~\cite{Kawasaki:1994,Dean:1996},
\begin{equation}
    \label{eq:DeanKawasaki}
    \partial_t \rhom = \partial_x \left[ D_0 \partial_x \rhom
    +  \rhom (V' \star \rhom)
    + \sqrt{2 D_0 \rhom} \: \eta \right]
    \:,
\end{equation}
with $(V' \star \rhom)(x) = \int V'(x-y) \rhom(y) \dd y$ and $\eta(x,t)$ a Gaussian white noise with $\moy{\eta(x,t)\eta(x',t')} = \delta(x-x')\delta(t-t')$. It is an exact microscopic equation for the density, which is equivalent to the set of Langevin equations~\eqref{eq:EqBrownianPart}. This equation plays an important role, as it has served as a starting point to many works over the last decades, see Ref.~\cite{Illien:2024} for a recent review.
Indeed, in principle Eq.~\eqref{eq:DeanKawasaki} gives access to the microscopic density field $\rhom(x,t)$ and thus to many observables.
However, because of the non locality induced by the interaction potential and the presence of the multiplicative noise, this equation remains mostly formal. In practice, explicit solutions are obtained from Eq.~\eqref{eq:DeanKawasaki} in limiting situations such as weak potentials, for which a linearized DKE has been derived~\cite{Demery:2014,Dean:2014,Demery2016,Poncet2017universal,Kruger2017,Kruger2018,Mahdisoltani:2021,venturelli2024,muzzeddu2024}.

In parallel, the fluctuating hydrodynamics~\cite{Spohn:1983} and the macroscopic fluctuation theory (MFT, not to be confused with mean field theory)~\cite{Bertini:2015} provide a systematic approach for determining the behavior of the density field at scale $\Nsc$ (a large adimensional rescaling factor, e.g.~proportional to system size), which obeys another stochastic partial differential equation~\cite{Spohn:1983,Bertini:2015,Olivencia:2017},
\begin{equation}
    \label{eq:flucHydro}
   \partial_t \rho = \partial_x \left[
    D(\rho) \partial_x \rho
    + \sqrt{\frac{\sigma(\rho)}{\Nsc} } \: \eta
    \right]
    \:,
\end{equation}
where $D(\rho)$ is the collective diffusion coefficient, $\sigma(\rho)$ the (collective) mobility, and $\eta(x,t)$ is still a Gaussian white noise. The density $\rho(x,t) = \rhom(\Nsc x, \Nsc^2 t)$
results from a coarse-graining of the microscopic density $\rhom$ involved in~\eqref{eq:DeanKawasaki}, and now describes the system on large scales. Note that the scalings of space and time are related, which is reminiscent of the fact that Eq.~\eqref{eq:flucHydro} holds for potentials $V$ which lead to a diffusive scaling for the density~\footnote{In practice, this corresponds to potentials for which $\int_{-\infty}^\infty (\e^{-\beta V(x)}-1)\dd x < \infty$, as implicitly assumed in this Letter. This defines what we call ``general'' potentials.}.

We stress that the evolution equations~(\ref{eq:DeanKawasaki},\ref{eq:flucHydro}) are very different: (i) Eq.~\eqref{eq:DeanKawasaki} describes the system at the microscopic scale, while~\eqref{eq:flucHydro} gives a coarse-grained description; (ii) contrary to Eq.~\eqref{eq:DeanKawasaki}, the macroscopic equation~\eqref{eq:flucHydro} is local, with all microscopic details encoded in the transport coefficients $D(\rho)$ and $\sigma(\rho)$; (iii) the noise term in the macroscopic equation~\eqref{eq:flucHydro} is inherently small due to coarse-graining ($\Nsc \gg 1$), enabling well-controlled weak-noise approaches at any temperature --- unlike Eq.~\eqref{eq:DeanKawasaki}, where such methods are limited to low temperatures.
Note that, recently, the DKE has been studied within a Martin-Siggia-Rose path integral formalism~\cite{Martin:1973} combined with a saddle-point analysis~\cite{Dandekar:2023,Illien:2025}. While Ref.~\cite{Dandekar:2023} addresses also the case of long-range interactions not covered by our approach, see~\cite{Note1}, the methodology used lacks a clear large parameter to systematically control the expansion (while the MFT naturally
features
$\Lambda \gg 1$). In turn, this affects the quantitative accuracy of some results [see the Supplemental Material (SM)~\cite{SM} for a comparison with the results of this Letter].

In turn, this last point has rendered the MFT formalism fully operational, and has permitted to obtain many explicit results for a wide class of diffusive systems, such as lattice gases, mass transfer models or hard-core Brownian particles~\cite{Derrida:2009a,Krapivsky:2012,Krapivsky:2014,Meerson:2014,Krapivsky:2015,Krapivsky:2015a,Sadhu:2015,Sadhu:2016,Zarfaty:2016,Derrida:2019b,Bettelheim:2022,Bettelheim:2022a,Mallick:2022,Krajenbrink:2022,Hurtado:2014,Espigares:2016,Escamilla:2017a,Escamilla:2017}.
However, surprisingly, the MFT has essentially not been applied to determine correlations in the key model of Brownian particles with general pairwise interaction potential described by~\eqref{eq:EqBrownianPart}~\footnote{The only exception is~\cite{Poncet:2021}, but it was restricted to the determination of the first-order correlation profile, corresponding to the large-scale behavior of the term $n=1$ in~\eqref{eq:DefCorrProf}, and only for the specific case of annealed initial conditions.}.
This is the object of this Letter.
More precisely, by combining the MFT formalism, including recent results obtained in this field as well as new ones derived in this Letter, with equilibrium statistical mechanics, we show that exact expressions can be obtained for the correlations and related key observables.
We finally show how the MFT formalism can be extended in higher dimensions to provide explicit expressions for correlation functions. Our approach bypasses the limitations of the DKE~\eqref{eq:DeanKawasaki} and provides a unified framework for studying out-of-equilibrium phenomena in interacting systems.

\emph{Transport coefficients for pairwise interacting Brownian particles.---} The first step to make~\eqref{eq:flucHydro} operational consists in the determination of $D(\rho)$ and $\sigma(\rho)$. From Eq.~\eqref{eq:flucHydro}, the collective diffusion coefficient $D(\rho)$ quantifies the linear response of the mean current of particles to the application of a small density gradient, while $\sigma(\rho)$ quantifies the fluctuations of current in equilibrium. Due to fluctuation-dissipation arguments, it is well known that $2 k_B T D(\rho) = f''(\rho) \sigma(\rho)$, where $f(\rho)$ is the equilibrium free energy density~\cite{Spohn:1991,Derrida:2007,Bertini:2015,Derrida:2025a}.
Furthermore, the mobility $\sigma$ can also be defined as the linear response of the mean current $\moy{j}$ to the application of an external force $F_0$ on all the particles, $\moy{j} = \sigma(\rho) F_0/(2 k_B T)$. Additionally, the effect of the constant external force on the dynamics~\eqref{eq:EqBrownianPart} of Brownian particles can be absorbed by a change of reference frame, with velocity $v = \mu_0 F_0$. Thus, in the fixed reference frame one has $\moy{j} = \rho v = \rho \mu_0 F_0$. Combining these relations yields
\begin{equation}
    \label{eq:TrCoefsExplicit}
    \sigma(\rho) = 2  \mu_0
    k_{\mathrm{B}} T \rho
    \quad \text{and} \quad
    D(\rho) = \mu_0 \: \partial_\rho P(\rho)
    \:,
\end{equation}
where we have introduced the equilibrium pressure $P(\rho)$ which satisfies $\partial_\rho P(\rho) = \rho f''(\rho)$~\footnote{Note that
$P(\rho)$ is the classical pressure for a system of particles of mass $m$ interacting pairwise via the potential $V$. The mass does not enter the expression of the pressure $P(\rho)$, see SM~\cite{SM} for more details.}. These arguments directly recover the known \textit{but unused in MFT} forms of $D(\rho)$ and $\sigma(\rho)$~\cite{Lekkerkerker:1981,Cichocki:1991,Butta:1999,Felderhof:2009}.
However, the practical use of equations~(\ref{eq:flucHydro},\ref{eq:TrCoefsExplicit}) requires the knowledge of the pressure at equilibrium, i.e.~the equation of state of the system. In the following, we will obtain fully explicit results by using either (i) the important specific choice of the interaction potential $V(x) = a/x^2$, corresponding to the Calogero model~\cite{Calogero:1975,Moser:1975} which is involved in many studies~\cite{Polychronakos:2006,Touzo:2024,Touzo:2024a}, for which the pressure can be computed explicitly~\cite{Choquard:2000,Lewin:2022}, see SM~\cite{SM},
or (ii) a general potential $V$,
for which various approximations have been developed to estimate the pressure, such as the Percus–Yevick approximation or hypernetted chain~\cite{Hansen:2005}. Here, we will use the virial expansion~\cite{Hill:1986} to compare our results to numerical simulations, which combined with~\eqref{eq:TrCoefsExplicit} gives
\begin{align}
    \label{eq:DfromVirial}
    &\frac{D(\rho)}{\mu_0 k_B T}
     \:=\:
    1 - \rho \int_{-\infty}^\infty \fm(x) \dd x
    \\
    &- \rho^2 \int_{-\infty}^\infty \dd x
    \int_{-\infty}^{\infty}\dd y \fm(x) \fm(y) \fm(x-y)
    + O( \rho^3 )
    \:, \nonumber
\end{align}
where $\fm(x) = \e^{-\beta V(x)} - 1$ is the Mayer function and $\beta = 1/(k_B T)$. Now that we are equipped with fully explicit expressions of $D(\rho)$ and $\sigma(\rho)$, we show that the MFT formalism allows
to obtain exact analytical expressions for correlation functions. While a variety of observables can be obtained, for the sake of clarity, we focus here on the paradigmatic case of a tracer particle immersed in a 1D system, and analyse its correlations with the bath of surrounding particles.

\emph{Bath-tracer correlations.---} The traditional observable in this case is the position $X_t$ of the tracer, which can be seen as one of the Brownian particles of the dynamics~\eqref{eq:EqBrownianPart}, with initial condition $X_0 = 0$.
The single-file constraint that the particles cannot bypass each other leads to a subdiffusive behavior $\moy{X_t^2} \sim \sqrt{t}$, which has been the focus of many theoretical and experimental studies~\cite{Harris:1965,Levitt:1973,Arratia:1983,Hahn:1996,Wei:2000,Kollmann:2003,Lutz:2004b,Lin:2005,Barkai:2009,Leibovich:2013,Hegde:2014,Sadhu:2015,Imamura:2017,Imamura:2021}.
This anomalous behavior results from the strong correlations present in the 1D geometry, which are quantified by the bath-tracer correlation profiles $\moy{\rhom(X_t+x,t) X_t^n}_c$ in the reference frame of the tracer, defined by their generating function
\begin{equation}
    \label{eq:DefCorrProf}
    \frac{\moy{\rhom(X_t + x,t) \e^{\lambda X_t}}}{\moy{\e^{\lambda X_t}}}
    = \sum_{n=0}^\infty \frac{\lambda^n}{n!} \moy{\rhom(X_t + x,t) X_t^n}_c
    \:,
\end{equation}
where $\moy{\cdot}_c$ denotes the connected correlations --- for instance, $\moy{\rhom(X_t+x,t) X_t}_c = \moy{\rhom(X_t+x,t) X_t} - \moy{\rhom(X_t+x,t)}\moy{X_t}$ --- and $\moy{\cdot}$ denotes the average over both the stochastic dynamics and the random initial condition (picked from the equilibrium distribution).
The importance of these profiles has been discussed in~\cite{Poncet:2021,Grabsch:2023a}, and they have been recently determined for emblematic exclusion models~\cite{Grabsch:2022,Grabsch:2023}. Actually, these are fundamental observables for the following reasons: (i) the $x$-dependence of these correlations gives access to the spatial structure of the response of the bath to the perturbation induced by the motion of the tracer; (ii) it can be shown (see below) that these profiles are non stationary, with a scaling behavior
\begin{equation}
    \label{eq:ProfAsympt}
    \frac{\moy{\rhom(X_t + x,t) \e^{\lambda X_t}}}{\moy{\e^{\lambda X_t}}}
    \underset{t \to \infty}{\simeq}
    \Phi \left( z \equiv \frac{x}{\sqrt{t}} ; \lambda \right)
    \:,
\end{equation}
which reflects their intrinsically dynamical nature, associated to the diffusive $\sqrt{t}$ growth of the perturbation of the bath; we stress that in particular, they are not encoded in the Gibbs equilibrium measure, as opposed to equilibrium pair correlation functions; (iii) in turn, the way the bath particles readjust at the front and behind the tracer controls its displacement; quantitatively, it has been found very recently that the correlation profiles $\Phi(z;\lambda)$ determine all the cumulants of $X_t$~\cite{Grabsch:2024b}, since
\begin{equation}
    \label{eq:CumulFromPhi}
    \frac{1}{\sqrt{t}} \ln \moy{\e^{\lambda X_t}}
    \underset{t \to \infty}{\simeq}
    -2 \frac{\partial_z P(\Phi)}{k_B T \Phi} \Bigg|_{0}
    \int_{\Phi(0^-)}^{\Phi(0^+)} D(r) \dd r
    \:.
\end{equation}
Finally, the large-scale correlation profile $\Phi$ is a fundamental observable, which we now determine with the help of the macroscopic equation~\eqref{eq:flucHydro} and the transport coefficients~\eqref{eq:TrCoefsExplicit}.

Since we aim to describe the system at a macroscopic time $t$, we choose the rescaling factor involved in~\eqref{eq:flucHydro} as $\Nsc = \sqrt{t/t_0}$, with $t_0$ a fixed time scale which we will show to be irrelevant. A key feature of the single-file geometry is that the position $X_t$ of the tracer is fully determined by the density $\rho(x,t)$~\cite{Krapivsky:2014,Krapivsky:2015a}. Indeed, since the particles cannot cross, the number of particles to the right of the tracer is conserved, hence we have that $X_t = \Nsc Y[\rho]$, where $Y[\rho]$ is defined by $\int_0^{Y[\rho]} \rho(x,t_0) \dd x = \int_0^\infty [\rho(x,t_0) - \rho(x,0)]\dd x$ (see SM~\cite{SM}).
This allows to analyse the statistics of $X_t$ from the evolution equation~\eqref{eq:flucHydro}, completed by an initial condition.
Essentially two types of initial conditions have been considered in the MFT literature: annealed initial conditions, for which $\rho(x,0)$ is random and corresponds to a system at equilibrium with mean density $\bar\rho$, and quenched initial conditions, for which $\rho(x,0) = \bar\rho$ is fixed.
Here,
since
we have access to the temperature dependence of the transport coefficients through~\eqref{eq:TrCoefsExplicit}, as opposed to most models considered in MFT which are athermal exclusion processes, we consider a more general class of initial conditions where the system is initially ($t=0$) at equilibrium at temperature $T_i$, different from the temperature $T$ defining the dynamics at $t>0$. This choice allows to describe the relevant case of a sudden quench of temperature~\cite{Rohwer:2017,Khalilian:2019}. The distribution of the initial density takes the form $P_0[\rho(x,0)] \asymp \e^{-\Nsc F[\rho(x,0)]}$, where the symbol $\asymp$ means that $\Lambda^{-1} \ln P_0[\rho(x,0)] \underset{\Lambda \to \infty}{\simeq} -F[\rho(x,0)]$, with~\cite{Derrida:2007,Derrida:2009a,Bertini:2015}
\begin{multline}
    \label{eq:InitDistr}
    F[\rho(x,0)] = \frac{1}{k_B T_i} \int \dd x [
    f_i(\rho(x,0)) - f_i(\bar\rho)
    \\
    - f_i'(\bar\rho) (\rho(x,0) - \bar\rho)
    ]
    \:,
\end{multline}
and $f_i(\rho)$ is the free energy density for a system at density $\rho$ and temperature $T_i$.

For the sake of consistency, and because the case of a sudden quench of temperature has not been considered in the MFT literature, we now recall the main steps underlying the MFT approach~\cite{Bertini:2015}. From the evolution equation~\eqref{eq:flucHydro}, one constructs a path integral representation of the probability of observing a given time evolution $P[\{ \rho(x,\tau) \}_{0\leq \tau \leq t_0}] \asymp \int \mathcal{D}H \e^{-\Nsc S[\rho,H]}$, with $S$ the MFT action and $H$ a conjugate field that imposes the conservation of the number of particles, which can be seen as a chemical potential that generates fluctuations of $\rho$ different from its typical evolution. This allows to write for instance $\moy{ \e^{\lambda X_t} } \asymp \int \mathcal{D}\rho \int \mathcal{D}H \e^{-\Nsc (S[\rho,H] - \lambda Y[\rho] + F[\rho(x,0)])}$. Since we consider a macroscopic density on a scale $\Nsc \gg 1$, this integral is dominated by the optimal evolution $(q,p)$ of $(\rho,H)$, which minimises the action $S + F - \lambda Y$. This is an important simplification that arises \textit{only} in the macroscopic formalism, and \textit{not} for the microscopic equation~\eqref{eq:DeanKawasaki}. The optimal evolution $(q,p)$ obeys the Euler-Lagrange equations, which give~\cite{Derrida:2009a,Krapivsky:2014,Krapivsky:2015a}
\begin{subequations}
  \label{eq:MFTbulk}
  \begin{align}
    \partial_\tau q
    &= \partial_x \left[
      D(q) \partial_x q - \sigma(q) \partial_x p
      \right]
      \:,
    \\
    \partial_\tau p
    &= - D(q)\partial_x^2 p - \frac{1}{2} \sigma'(q) (\partial_x p)^2
      \:,
  \end{align}
\end{subequations}
for $0 \leq \tau \leq t_0$, and the boundary conditions~\footnote{Note that there are subtleties in the treatment of these conditions since $q(x,t_0)$ can be discontinuous at $x=Y[q]$~\cite{Krapivsky:2014,Krapivsky:2015a}. Since we focus here on the first order in $\lambda$ only, this is not a problem. To study higher orders, other approaches can be used, see for instance~\cite{Imamura:2017,Imamura:2021,Grabsch:2024b}.}
\begin{gather}
    \label{eq:MFTfinal}
    p(x,t_0) = \frac{\lambda}{q(Y[q],t_0)} \Theta(x - Y[q])
    \:, \\
    \label{eq:MFTinit}
    p(x,0) = \frac{\lambda}{q(Y[q],t_0)} \Theta(x)
    + \int_{\bar\rho}^{q(x,0)} \frac{2D_i(r)}{\sigma_i(r)} \dd r
    \:,
\end{gather}
where $D_i(\rho)$ and $\sigma_i(\rho)$ denote the transport coefficients~\eqref{eq:TrCoefsExplicit} at temperature $T_i$, $\Theta$ is the Heaviside step function which arises from the study of $X_t$, and the last term in~\eqref{eq:MFTinit} comes from the equilibrium initial distribution~\eqref{eq:InitDistr}.
The large-scale behavior of the correlation profiles~\eqref{eq:DefCorrProf} is obtained similarly to $\moy{\e^{\lambda X_t}}$. Indeed, noticing that the numerator and the denominator in~\eqref{eq:DefCorrProf} involve the same exponential factor, and thus the same optimal evolution,
\begin{equation}
    \label{eq:ResultCorrProf}
    \frac{\moy{\rhom(X_t + x,t) \e^{\lambda X_t}}}{\moy{\e^{\lambda X_t}}}
    \underset{\Nsc \to \infty}{\simeq}
    q \left( Y[q] + \frac{x}{\Nsc}, t_0 \right)
    \equiv \Phi(z;\lambda)
    \:,
\end{equation}
which is indeed a function of $z = x/\sqrt{t}$ since $\Nsc = \sqrt{t/t_0}$~\footnote{Note that the limit $t \to \infty$ is thus equivalent to $\Lambda \to \infty$, which holds for $\Lambda \gg 1$.}.

\begin{figure*}
    \centering
    \includegraphics[width=0.27\textwidth]{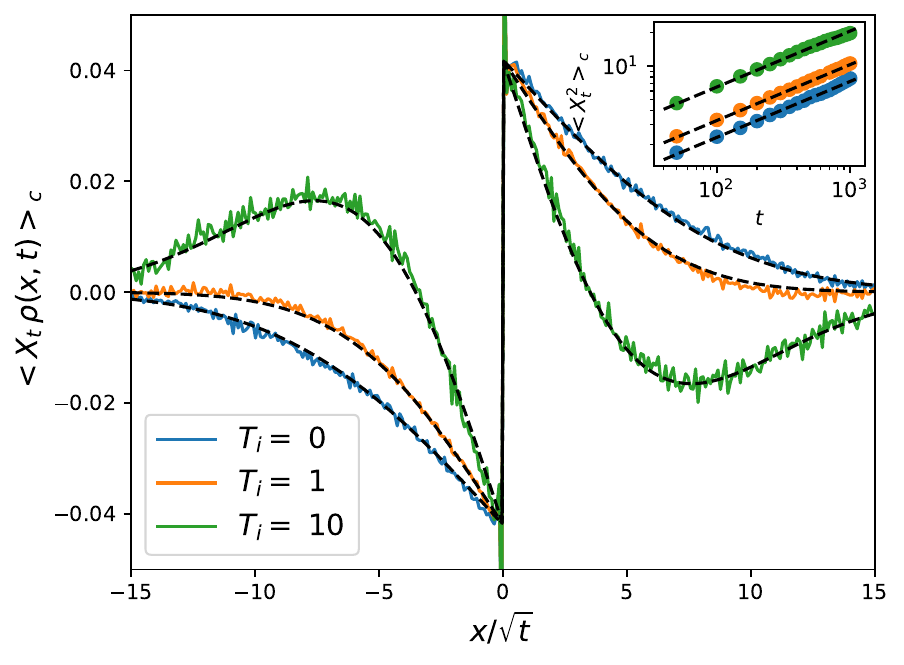}
    \includegraphics[width=0.27\textwidth]{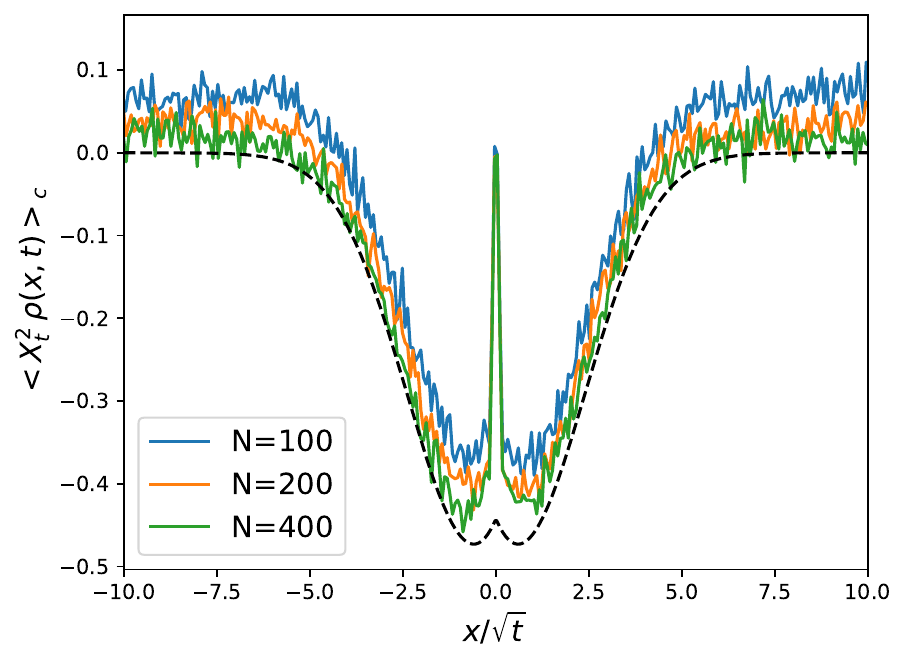}
    \includegraphics[width=0.27\textwidth]{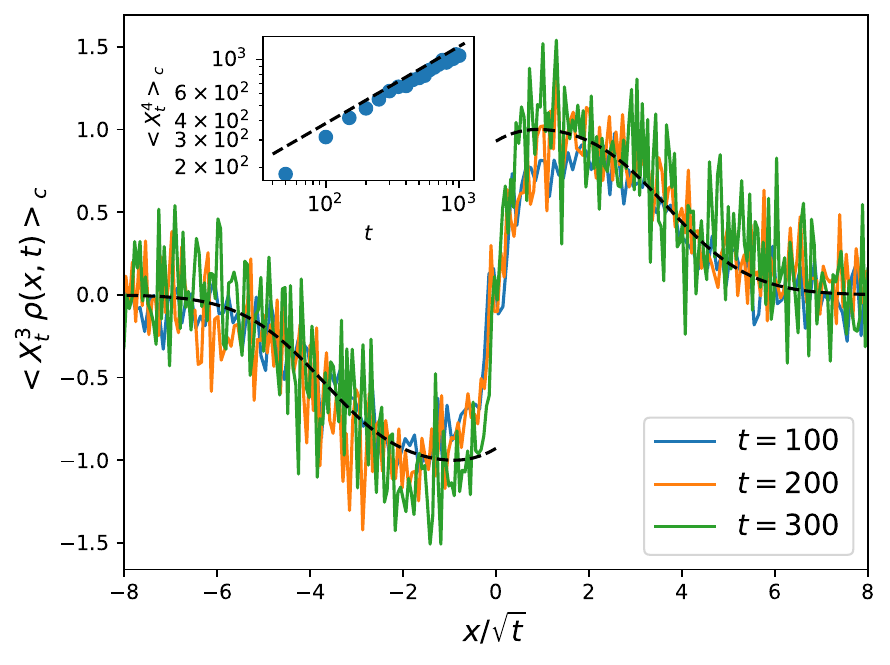}
    \caption{Bath-tracer correlation profile $\moy{X_t^n \rho(X_t+x,t)}_c$ for the Calogero potential $V(x) = 1/x^2$ at temperature $T=1$, with $\mu_0 = k_B = 1$. Insets: cumulants $\moy{X_t^n}_c$ as a function of $t$ in log-log scale. The solid lines are obtained from numerical simulations (with $N = 400$ particles and $10^4$ realisations), while the dashed lines correspond to the theoretical predictions~\eqref{eq:Phi1} with~\eqref{eq:TrCoefsExplicit}. Left: $n=1$ at $t=100$, with $\bar\rho = 1$ and different initial temperatures $T_i$. Middle: $n=2$ with $\bar\rho = 0.2$ and $T_i = T$. We illustrate the convergence to our prediction as $N$ is increased. Right: $n=3$ with $\bar\rho = 0.2$ and $T_i = T$.}
    \label{fig:ProfCalo}
\end{figure*}

\begin{figure*}
    \centering
    \includegraphics[width=0.27\textwidth]{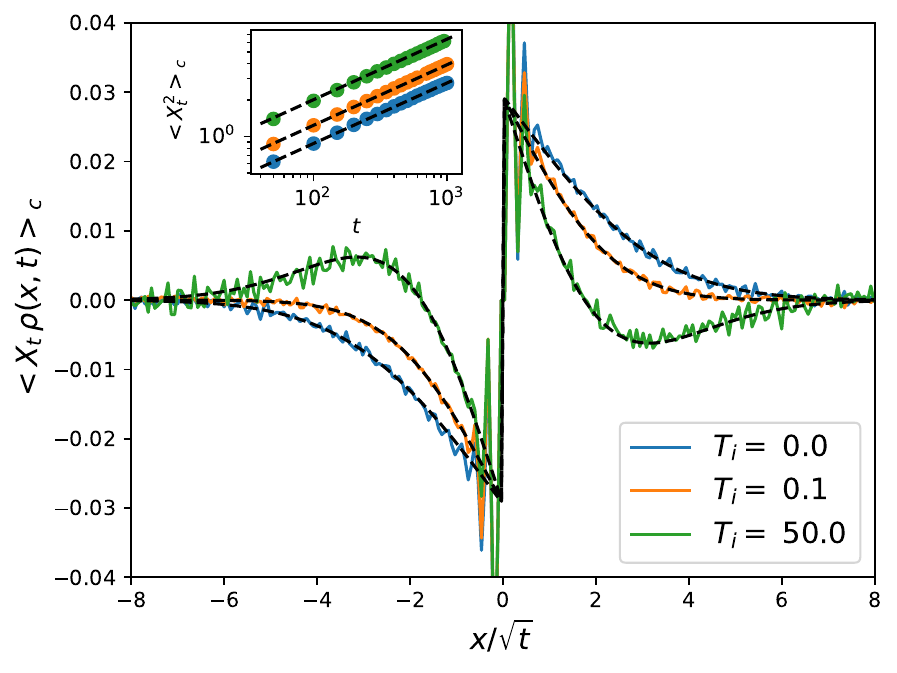}
    \includegraphics[width=0.27\textwidth]{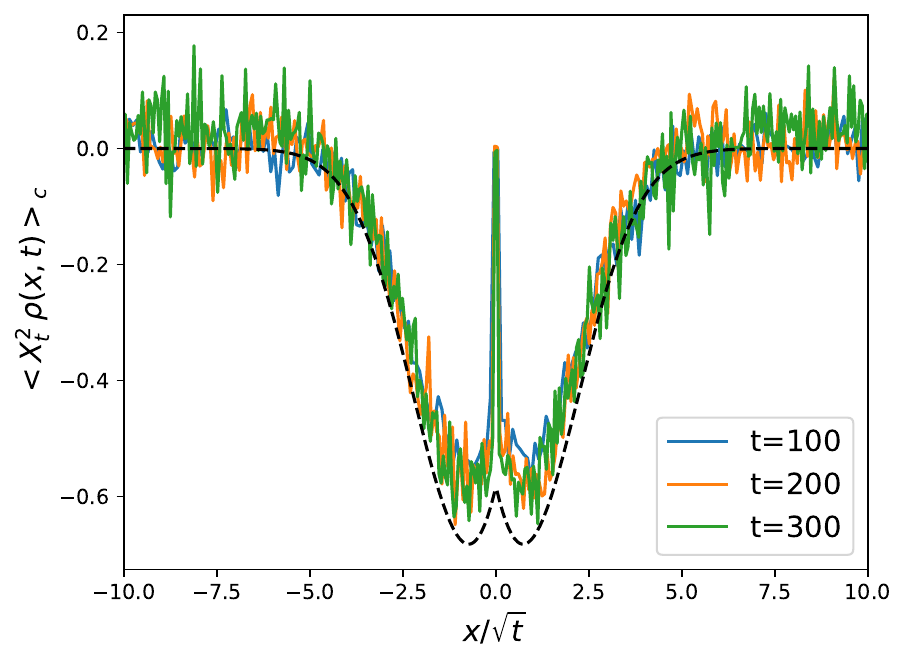}
    \includegraphics[width=0.27\textwidth]{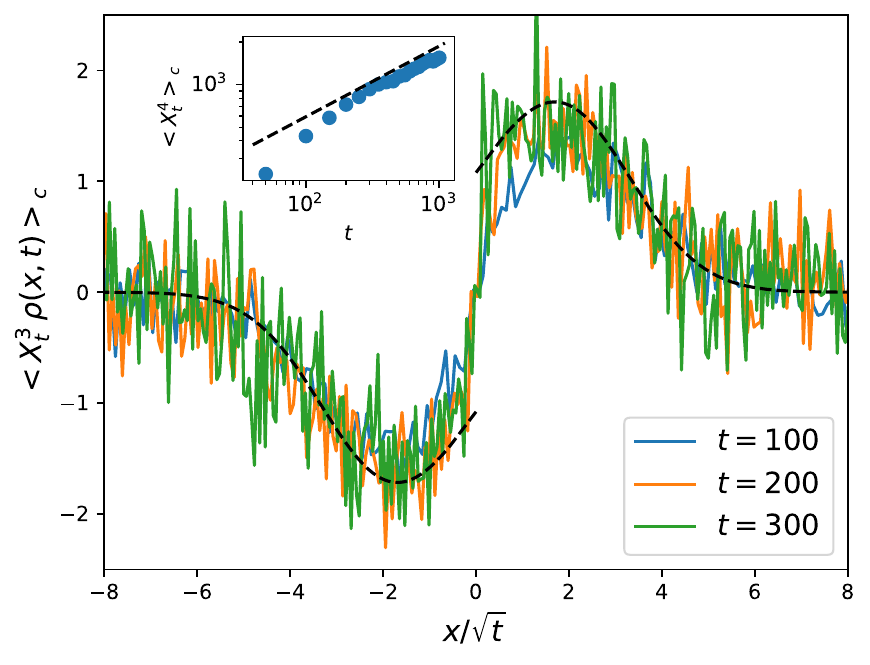}
    \caption{Bath-tracer correlation profile $\moy{X_t^n \rho(X_t+x,t)}_c$ for the Weeks-Chandler-Andersen potential ($V(x) = 4 (x^{-12} - x^{-6}) + 1$ for $\abs{x}<2^{1/6}$ and $0$ otherwise) at temperature $T_i = T =1$ with $\mu_0 = k_B = 1$. Insets: cumulants $\moy{X_t^n}_c$ as a function of $t$ in log-log scale. The solid lines are obtained from numerical simulations (with $N = 400$ particles and $10^4$ realisations), while the dashed lines correspond to the theoretical predictions~\eqref{eq:Phi1} with~\eqref{eq:TrCoefsExplicit}. Left: $n=1$ at $t=100$ with $\bar\rho = 0.7$, $T=0.1$, and different values of $T_i$. For these values of $T$ and $\bar\rho$, the virial expansion~\eqref{eq:DfromVirial} is not a good approximation. We thus determined $D(\bar\rho) \simeq 1.7$ by fitting the data for $T_i = T$ using~\eqref{eq:Phi1}. The obtained value is then used for the other curves. Middle: $n=2$ with $\bar\rho = 0.2$ and $T_i = T = 1$. Right: $n=3$ with $\bar\rho = 0.2$ and $T_i = T = 1$. In these last two plots, the diffusion coefficient is computed using the virial expansion~\eqref{eq:DfromVirial}.}
    \label{fig:ProfWCA}
\end{figure*}

No solution of the MFT equations~(\ref{eq:MFTbulk},\ref{eq:MFTfinal},\ref{eq:MFTinit}) is currently available for general $D(\rho)$ and $\sigma(\rho)$. Exact solutions have recently been obtained for specific $D(\rho)$ and $\sigma(\rho)$~\cite{Grabsch:2022,Mallick:2022,Bettelheim:2022,Bettelheim:2022a,Krajenbrink:2022,Grabsch:2024}, which do not include the case~\eqref{eq:TrCoefsExplicit} studied here. Nevertheless, the first orders in $\lambda$ can be computed explicitly to obtain exactly the first correlation profiles $\moy{\rho_0(x,t) X_t^n}_c$~\eqref{eq:DefCorrProf}. Indeed, writing $q = \bar\rho + \lambda q_1 + O(\lambda^2)$, $p = \lambda p_1 + O(\lambda^2)$ and $Y[q] = \lambda Y_1 + O(\lambda^2)$,
Eqs.~\eqref{eq:MFTbulk} reduce to diffusion equations at first order in $\lambda$, which can be solved to finally give from~\eqref{eq:ResultCorrProf},
\begin{multline}
    \label{eq:Phi1}
    \Phi_1(z) \equiv q_1 \left( \frac{x}{\Nsc}, t_0 \right) = \frac{\sigma_i(\bar\rho)}{4  \bar\rho D_i(\bar\rho)}
    \erfc \left( \frac{z}{\sqrt{4 D(\bar\rho)}} \right)
    \\
    + \left( \frac{\sigma(\bar\rho)}{4 \bar\rho D(\bar\rho)}
    - \frac{\sigma_i(\bar\rho)}{4 \bar\rho D_i(\bar\rho)}
    \right)
    \erfc \left( \frac{z}{\sqrt{8 D(\bar\rho)}} \right)
    \:,
\end{multline}
for $z>0$ and $\Phi_1(-z) = -\Phi_1(z)$. Here we have denoted $\Phi_1$ the first order in $\lambda$ of~\eqref{eq:ResultCorrProf}, which corresponds to the large-scale behavior of $\moy{X_t \rho_0(X_t+x,t)}_c$. Note that this expression does not depend on the arbitrarily chosen time scale $t_0$, as claimed above.
Several comments are in order.
(i) The profile $\Phi_1$ is especially important since it provides the large-scale behavior of the covariance $\mathrm{Cov}(X_t, \rho(X_t+x))$, and thus quantifies the correlations between the displacement of the tracer, and the density of bath particles.
(ii) When $T_i = T$, we recover the correlations of the usual annealed case~\cite{Poncet:2021}, where only the term $\propto \erfc ( \frac{z}{\sqrt{4 D(\bar\rho)}} )$ remains;
the other usual situation of a quenched case is shown in the SM~\cite{SM} to involve only the term $\propto \erfc ( \frac{z}{\sqrt{8 D(\bar\rho)}} )$; in both situations, $\Phi_1(x)$ is positive for $x>0$ (and negative for $x<0$), indicating that an increase of $X_t$ is associated with an increase of the density in front of the tracer (and a decrease behind).
(iii) The expression~\eqref{eq:Phi1} allows us to unveil a surprising effect in the physically relevant situation discussed above of a sudden quench of temperature $T_i \neq T$, in which both terms in~\eqref{eq:Phi1} remain: as soon as $T_i/D_i(\bar\rho) > T/D(\bar\rho)$ ---  which is generally the case if $T_i>T$ and $V$ is purely repulsive --- the profile $\Phi_1$ becomes non-monotonic and even changes sign with the distance (see Figs.~\ref{fig:ProfCalo} and~\ref{fig:ProfWCA} left): in other words, the displacement $X_t$ induces a nontrivial spatial structure of the bath particles, with first an accumulation of particles in front of the tracer, and then a depletion further away (and an opposite behaviour behind). Note that this effect is nonstationary and occurs on large scales $\sim \sqrt{t}$.
(iv) The determination of the profile~\eqref{eq:Phi1} directly yields the expression of the mean squared displacement from~\eqref{eq:CumulFromPhi},
\begin{equation}
    \frac{\moy{X_t^2}_c}{\sqrt{t}}
    \underset{t \to \infty}{\simeq}
    \frac{\sigma_i(\bar\rho)}{\bar\rho^2 D_i(\bar\rho)}
    \sqrt{\frac{D(\bar\rho)}{\pi}}
    + \frac{D_i(\bar\rho) \sigma(\bar\rho) - D(\bar\rho) \sigma_i(\bar\rho)}
    {\bar\rho^2 D_i(\bar\rho) \sqrt{2\pi D(\bar\rho)} }
    \:.
\end{equation}
In the case $T_i = T$, we recover the
well-known
result of Kollmann~\cite{Kollmann:2003}. In the case $T_i/D_i(\bar\rho) > T/D(\bar\rho)$ for which $\Phi_1$ is non-monotonic, the fluctuations of $X_t$ are larger than in the annealed case $T_i = T$, indicating that the temperature quench increases the fluctuations.

\emph{Higher-order bath-tracer correlations.---} An important aspect of the MFT formalism is that, in contrast with the linearized DKE, it allows to go beyond the Gaussian fluctuations. As an illustration, we consider the higher-order correlation profiles $\Phi_2$ and $\Phi_3$, and the corresponding $\moy{X_t^4}_c$ which probes the deviation from Gaussian behavior. For simplicity, we focus on the annealed initial condition $T_i = T$, and combine the expressions of these quantities obtained recently from MFT~\cite{Grabsch:2024b} (recalled in SM~\cite{SM}) with those of $D(\rho)$ and $\sigma(\rho)$~\eqref{eq:TrCoefsExplicit}.
Quantitative agreement with numerical simulations is displayed in Figs.~\ref{fig:ProfCalo} and~\ref{fig:ProfWCA} for the Calogero and Weeks-Chandler-Andersen (WCA) potentials~\footnote{The discrepancies observed near $x=0$ are due to microscopic effects, which are not captured by the MFT and which vanish for $t\to \infty$. The small deviation of the fourth cumulant from the prediction is due to finite size effects, see SM~\cite{SM}}.
We stress that our results are exact for any interaction potential
which yields a ``standard'' thermodynamics (e.g.~extensive energy)~\cite{Note1} and enforces the single-file constraint. The MFT formalism and the transport coefficients~\eqref{eq:TrCoefsExplicit} nevertheless still apply if the particles can bypass each other, as illustrated in the SM~\cite{SM} for a Gaussian interaction potential.

\emph{Higher-dimensional systems.---} The MFT formalism is not restricted to one-dimensional systems, and as an illustration of our approach, we now consider the dynamical density-density correlations.
These observables, defined in the absence of a tracer, are key theoretical and experimental quantities~\cite{Hansen:2005,Doi:1988,Stribeck:2007,Sadhu:2016}. Relying on the $d$-dimensional counterparts of the evolution equation~\eqref{eq:flucHydro} and initial distribution~\eqref{eq:InitDistr}, and treating the noise term as a perturbation (which is justified in the macroscopic description), we obtain (see SM~\cite{SM})
\begin{equation}
    \label{eq:DensDensCorrel}
    \moy{\rho_0(\boldsymbol{x},t) \rho_0(\boldsymbol{y},0)}_c
    \underset{\Lambda \to \infty}{\simeq}
    \frac{\sigma_i(\bar\rho)}{2 D_i(\bar\rho)}
    \frac{\e^{- \frac{(\boldsymbol{x}-\boldsymbol{y})^2}{4 D(\bar\rho) t}}}{(4\pi D(\bar\rho) t)^{d/2}}
    \:.
\end{equation}
Note that consequently the dynamical structure factor
$S(\boldsymbol{k},t) \equiv \int \e^{\I \boldsymbol{k} \cdot \boldsymbol{x}} \moy{\rho(\boldsymbol{x},t) \rho(\boldsymbol{0},0)}_c  \dd \boldsymbol{x}/\bar\rho$ takes the form $S(\boldsymbol{k},t) = \sigma_i(\bar\rho) \e^{-\boldsymbol{k}^2 D(\bar\rho) t}/(2\bar\rho D_i(\bar\rho))$. In the particular case $d=1$ with annealed initial condition $T_i=T$, we recover the result of Kollmann~\cite{Kollmann:2003}. Note that the use of the linearized DKE would lead to a similar expression, but with $D(\rho)$ replaced by $D_0 + \mu_0 \rho \int_{-\infty}^\infty V(x) \dd x$,
corresponding to
the small-$V$ expansion of~\eqref{eq:DfromVirial} (if $V$ is integrable)~\cite{Demery:2014,Illien:2025}. In other words, the linearized DKE cannot account for the density-density correlations for the Calogero and WCA potentials considered here (which are not integrable), as determined by~\eqref{eq:DensDensCorrel} in our approach.
Finally, we note that the result~\eqref{eq:DensDensCorrel} is exact for any interaction potential~\cite{Note1} for which the dynamics remains diffusive at large scales.
The determination of the precise range of validity of~\eqref{eq:DensDensCorrel}, which is the range of validity of the MFT, is still an open question.

\emph{Conclusion.---} Combining the MFT formalism with explicit expressions for the transport coefficients~\eqref{eq:TrCoefsExplicit} deduced from equilibrium statistical mechanics, we obtained exact expressions of correlation functions for the paradigmatic model of interacting Brownian particles. We stress that these results hold for general interaction potentials~\cite{Note1}, and both for equilibrium and nonequilibrium situations.
Here, we have illustrated the nonequilibrium setting by considering a sudden quench of temperature, but other situations have been studied within MFT, such as a step initial condition which yields a transient behavior at all times~\cite{Derrida:2009a}.
In addition to providing explicit results, we unveiled an unexpected and generic non-monotonic behavior with the distance of the simplest bath-tracer correlation profiles after a temperature quench.

On the methodological side, we stress that while the DKE yields a formally
exact microscopic equation, the MFT is a systematic tool to obtain explicit and exact results for the dynamics of interacting particle systems at large scales.
In turn, we expect that the ongoing developments of MFT calculations for arbitrary $D(\rho)$ and $\sigma(\rho)$, such as those obtained recently for tracer or current statistics in $1$D~\cite{Meerson:2014,Krapivsky:2014,Krapivsky:2015a,Grabsch:2024b,Bodineau:2025,Saha:2025} or for systems with additional conservation laws~\cite{Gutierrez:2019,Agranov:2021,Agranov:2023}, will allow to obtain further explicit results for interacting particle systems.

\emph{Acknowledgements.---} We thank Pierre Illien and Alexis Poncet for constructive discussions on related topics.

\bibliographystyle{apsrev4-2}

%

\clearpage
\widetext

\let\addcontentsline\oldaddcontentsline

\begin{center}
  \begin{large}

    \textbf{
     Supplemental Material for\texorpdfstring{\\}{}Exact large-scale correlations in diffusive systems with general interactions:\texorpdfstring{\\}{} explicit characterisation without the Dean--Kawasaki equation
   }
  \end{large}
   \bigskip

   Aurélien Grabsch, Davide Venturelli and Olivier B\'enichou
\end{center}

\setcounter{equation}{0}
\makeatletter

\renewcommand{\theequation}{S\arabic{equation}}
\renewcommand{\thefigure}{S\arabic{figure}}

\renewcommand{\bibnumfmt}[1]{[S#1]}
\renewcommand{\citenumfont}[1]{S#1}

\setcounter{secnumdepth}{3}

\tableofcontents

\section{Macroscopic fluctuation theory for a tracer}

\subsection{Microscopic to macroscopic description}

We consider an ensemble of Brownian particles, interacting pairwise with potential $V$. Their dynamics obeys the set of overdamped Langevin equations
\begin{equation}
    \label{eq:EqBrownianPartSM}
    \dt{x_i}{t} = - \mu_0 \sum_{j \neq i} V'(x_i-x_j)
    + \sqrt{2 D_0} \: \eta_i
    \:,
\end{equation}
with $\eta_i$ a Gaussian white noise of unit variance, $\mu_0$ the bare mobility of a particle, and $D_0 = \mu_0 k_B T$ the bare diffusion coefficient.
This ensemble of particles can be described by a density
\begin{equation}
    \label{eq:DefRho0SM}
    \rho_0(x,t) = \sum_n \delta \big( x-x_n(t) \big)
    \:.
\end{equation}
This is a \textit{microscopic} density, whose knowledge is equivalent to the one of the ensemble of positions $\{ x_i(t) \}$. The macroscopic density $\rho(x,t)$ is obtained by performing a diffusive rescaling of time and space with a large parameter $\Nsc$,
\begin{equation}
    \label{eq:DefRhoSM}
    \rho(x,t) = \rho_0(\Nsc x, \Nsc^2 t)
    \:.
\end{equation}

Let us additionally consider a tracer, which we can choose to correspond to the particle with $i=0$, so that $X_t = x_0(t)$, with the convention that $X_0 = 0$. In one dimension, with sufficiently strong repulsive potential at short distance so that the particles cannot cross, we can determine the position of $X_t$ from the knowledge of the density $\rho_0$ (or $\rho$) since the number of particles to the right of $X_t$ is conserved, i.e.,
\begin{equation}
    \int_{X_t}^\infty \rho_0(x,t) \dd x
    = \int_0^\infty \rho_0(x,0) \dd x
    \:.
\end{equation}
However, this equation needs to be regularised, since both terms are formally infinite, so that~\cite{Krapivsky:2014SM,Krapivsky:2015aSM}
\begin{equation}
    \int_0^{X_t} \rho_0(x,t) \dd x
    = \int_0^\infty [\rho_0(x,t) - \rho_0(x,0)] \dd x
    \:.
\end{equation}
Performing the diffusive rescaling~\eqref{eq:DefRhoSM} this renders for the macroscopic density
\begin{equation}
    \label{eq:defXtfromRhoSM}
    \int_0^{X_{\Nsc^2 t}/\Nsc} \rho(x,t) \dd x
    = \int_0^\infty [\rho(x,t) - \rho(x,0)] \dd x
    \:.
\end{equation}
This is the relation used in the main text to define $X_{\Nsc^2 t} = \Nsc Y[\rho]$ from the macroscopic density $\rho(x,t)$.

\subsection{MFT action and optimal evolution}

The macroscopic density $\rho(x,\tau)$ obeys the evolution equation
\begin{equation}
    \label{eq:flucHydroSM}
   \partial_\tau \rho = \partial_x \left[
    D(\rho) \partial_x \rho
    + \sqrt{\frac{\sigma(\rho)}{\Nsc} } \: \eta
    \right]
    \:.
\end{equation}
We sketch here the main steps of the derivation of the MFT equations (10-12) of the main text. First, we formally write from~\eqref{eq:flucHydroSM} the probability of observing a given evolution from an initial profile $\rho(x,0)$ to a profile $\rho(x,t/\Nsc^2)$ at a given large microscopic time $t$, corresponding to the finite macroscopic time $t/\Lambda^2$ (we keep the large microscopic time $t$ explicitly here since we will use it below to define $\Lambda$),
\begin{equation}
    P[\rho(x,0) \to \rho(x,t/\Nsc^2)] \asymp
    \int \D \eta \: \delta \left[
     \partial_\tau \rho - \partial_x \left(
    D(\rho) \partial_x \rho
    + \sqrt{\frac{\sigma(\rho)}{\Nsc} } \: \eta
    \right)
    \right]
    \e^{- \int \dd x \int_0^{t/\Nsc^2} \dd \tau \frac{\eta(x,\tau)^2}{2}}
    \:,
\end{equation}
where we have used that $\eta$ is a Gaussian white noise and where $\asymp$ means logarithmically equivalent, that is,
\begin{equation}
    \lim_{\Lambda \to \infty} \frac{1}{\Lambda}  \ln P[\rho(x,0) \to \rho(x,t/\Nsc^2)]
    =
    \lim_{\Lambda \to \infty} \frac{1}{\Lambda} \ln
    \int \D \eta \: \delta \left[
     \partial_\tau \rho - \partial_x \left(
    D(\rho) \partial_x \rho
    + \sqrt{\frac{\sigma(\rho)}{\Nsc} } \: \eta
    \right)
    \right]
    \e^{- \int \dd x \int_0^{t/\Nsc^2} \dd \tau \frac{\eta(x,\tau)^2}{2}}
    \:.
\end{equation}
Representing the delta function as an integral,
\begin{equation}
    \delta \left[
     \partial_\tau \rho - \partial_x \left(
    D(\rho) \partial_x \rho
    + \sqrt{\frac{\sigma(\rho)}{\Nsc} } \: \eta
    \right)
    \right]
    = \int \D H \e^{-\Nsc \int H \left[\partial_\tau \rho - \partial_x \left(
    D(\rho) \partial_x \rho
    + \sqrt{\frac{\sigma(\rho)}{\Nsc} } \: \eta
    \right)\right]},
\end{equation}
we can perform the Gaussian integration over $\eta$,
\begin{equation}
    \label{eq:ProbEvolSM}
    P[ \{ \rho(x,\tau) \}_{0 \leq \tau \leq t}] \asymp
    \int \D H \e^{- \Nsc S[\rho,H]}
    \:,
\end{equation}
where we introduced the MFT action~\cite{Derrida:2009aSM}
\begin{equation}
    \label{eq:MFTactionSM}
    S[\rho,H] = \int_{-\infty}^\infty \dd x
    \int_0^{t/\Nsc^2} \dd \tau \left[
    H \partial_\tau \rho + D(\rho) \partial_x \rho \partial_x H
    - \frac{\sigma(\rho)}{2} (\partial_x H)^2
    \right]
    \:.
\end{equation}
Combining the evolution~\eqref{eq:ProbEvolSM} with the definition of $X_t$~\eqref{eq:defXtfromRhoSM}, we can write
\begin{equation}
    \label{eq:MoyElambdaXtSM}
    \moy{\e^{\lambda X_t}}
    = \int \D \rho \int \D H \e^{-\Nsc S[\rho,H] + \Nsc \lambda Y[\rho]}
    P[\rho(x,0)]
    \:,
\end{equation}
where $P[\rho(x,0)]$ is the probability of observing a given initial profile $\rho(x,0)$. We assume that the system is initially at equilibrium at temperature $T_i$, around a mean density $\bar\rho$, therefore~\cite{Derrida:2009aSM},
\begin{equation}
    \label{eq:FinitSM}
    P[\rho(x,0)] \asymp \e^{-\Nsc F[\rho(x,0)]}
    \:,
    \quad
    F[\rho(x,0)]
    =
    \frac{1}{k_B T_i}
    \int_{-\infty}^\infty \dd x \left[
    f_i(\rho(x,0)) - f_i(\bar\rho)
- (\rho(x,0) - \bar\rho) f_i'(\bar\rho)
    \right]
    \:,
\end{equation}
with $f_i(\rho)$ the free energy density for an infinite system at equilibrium at temperature $T_i$ and density $\rho$. Combined with~\eqref{eq:MoyElambdaXtSM}, this gives
\begin{equation}
    \moy{\e^{\lambda X_t}}
    \asymp \int \D \rho \int \D H \e^{-\Nsc (S[\rho,H] - \lambda Y[\rho] + F[\rho(x,0)])}
    \:.
\end{equation}
Since we are interested in the long-time limit, we choose the rescaling factor
\begin{equation}
    \label{eq:ScalingFactorSM}
    \Nsc = \sqrt{\frac{t}{t_0}}
    \:,
\end{equation}
with $t_0$ an arbitrary fixed time scale which will play no role later. This way, for $t\to \infty$, the integral is dominated by the optimal evolution $(q,p)$ of $(\rho,H)$,
\begin{equation}
    \moy{\e^{\lambda X_t}}
    \asymp \e^{-\Nsc (S[q,p] - \lambda Y[q] + F[q(x,0)])}
    \:.
\end{equation}
The optimal path can be computed by evaluating $S- \lambda Y + F$ at $(q + \delta \rho, p + \delta H)$, and requiring that the terms linear in $\delta \rho$ and $\delta H$ vanish. This yields the MFT equations for $0 \leq \tau \leq t_0$,
\begin{subequations}
  \label{eq:MFTbulkSM}
  \begin{align}
    \partial_\tau q
    &= \partial_x \left[
      D(q) \partial_x q - \sigma(q) \partial_x p
      \right]
      \:,
    \\
    \partial_\tau p
    &= - D(q)\partial_x^2 p - \frac{1}{2} \sigma'(q) (\partial_x p)^2
      \:,
  \end{align}
\end{subequations}
completed by the boundary conditions
\begin{equation}
    \label{eq:MFTfinalSM}
    p(x,t_0) = \frac{\lambda}{q(Y[q],t_0)} \Theta(x - Y[q])
    \:,
\end{equation}
\begin{equation}
    \label{eq:MFTinitSM}
    p(x,0) = \frac{\lambda}{q(Y[q],t_0)} \Theta(x)
    + \int_{\bar\rho}^{q(x,0)} \frac{2D_i(r)}{\sigma_i(r)} \dd r
    \:,
\end{equation}
where we have used that $f_i''(\rho) = 2 k_B T_i D_i(\rho)/\sigma_i(\rho)$ and~\cite{Krapivsky:2014SM,Krapivsky:2015aSM}
\begin{equation}
    \frac{\delta Y[\rho]}{\delta \rho(x,t_0)}
    = \frac{\Theta(x - Y[\rho])}{\rho(Y[\rho], t_0)}
    \:,
    \quad \text{and} \quad
    \frac{\delta Y[\rho]}{\delta \rho(x,0)}
    = -\frac{\Theta(x)}{\rho(Y[\rho], t_0)}
    \:.
\end{equation}
Similarly, using the rescaling~\eqref{eq:DefRhoSM} we have that
\begin{align}
    \moy{\rho_0(X_t + x, t) \e^{\lambda X_t}}
    &= \moy{ \rho(Y[\rho] + x/\Nsc, t/\Nsc^2) \e^{\lambda X_t}}
    \nonumber
    \\
    &= \int \D \rho \int \D H  \rho(Y[\rho] + x/\Nsc,t_0) \e^{-\Nsc (S[\rho,H] - \lambda Y[\rho] + F[\rho(x,0)])}
    \nonumber
    \\
    &\propto
    q(Y[q] + x/\Nsc,t_0) \e^{-\Nsc (S[q,p] - \lambda Y[q] + F[q(x,0)])}
    \:,
\end{align}
so that the generating function of the correlation profiles becomes
\begin{equation}
    \frac{\moy{\rho_0(X_t + x, t) \e^{\lambda X_t}}}{\moy{\e^{\lambda X_t}}}
    \underset{\Lambda \to \infty}{\simeq}
    q \left(Y[q] + x/\Nsc,t_0 \right)
    = q \left(Y[q] + \frac{x}{\sqrt{t}} \sqrt{t_0},t_0 \right)
    \equiv \Phi \left( z = \frac{x}{\sqrt{t}} ; \lambda \right)
    \:.
\end{equation}
We can actually show that $\Phi$ does not depend on the arbitrary scale $t_0$, by performing the change of functions
\begin{equation}
    \label{eq:ScalingTildeSM}
    \qt(x,\tau) = q \left( x \sqrt{t_0}, \tau t_0 \right)
    \:,
    \quad
    \pt(x,\tau) =  p \left( x \sqrt{t_0}, \tau t_0 \right)
    \:,
\end{equation}
which leaves invariant the evolution equations~\eqref{eq:MFTbulkSM}, the initial condition~\eqref{eq:MFTinitSM}, and only shifts the final condition~\eqref{eq:MFTfinalSM} to time $t_0 = 1$. Therefore, $\qt$ does not depend on $t_0$, thus
\begin{equation}
    \Phi(z;\lambda) = q(Y[q] + z \sqrt{t_0}, t_0)
    = \qt(Y[q]/\sqrt{t_0} + z , 1)
    \:.
\end{equation}
Finally, using~\eqref{eq:defXtfromRhoSM}, we have that
\begin{equation}
    Y[q] = Y[\tilde{q}] \sqrt{t_0},
\end{equation}
and thus
\begin{equation}
    \Phi(z;\lambda)
    = \qt(Y[\qt] + z , 1)
    \:,
\end{equation}
which is indeed independent of $t_0$. The change of functions~\eqref{eq:ScalingTildeSM} is usually performed in MFT so as to put the final time $t_0$ to $1$ without loss of generality~\cite{Derrida:2009aSM,Bertini:2015SM}. We have not done it in the main text, so as to keep the variables $x$ and $t$ with dimensions of space and time respectively, at the cost of introducing the irrelevant time scale $t_0$ in the MFT equations.

\subsection{Solution at first order in \texorpdfstring{$\lambda$}{lambda}}
\label{sec:MFTorder1}

The MFT equations~(\ref{eq:MFTbulkSM}-\ref{eq:MFTinitSM}) cannot be solved explicitly. However, they can be solved by expanding $q$ and $p$ in powers of $\lambda$ (this procedure was carried out in~\cite{Krapivsky:2014SM,Krapivsky:2015aSM} for the case $T_i = T$),
\begin{equation}
    \label{eq:ExpQPSM}
    q = \bar\rho + \lambda q_1 + \cdots
    \:,
    \quad
    p = \lambda p_1 + \cdots
    \:,
    \quad
    Y[q] = \lambda Y_1 + \cdots
\end{equation}
Inserting these expansions into~(\ref{eq:MFTbulkSM}-\ref{eq:MFTinitSM}), we get
\begin{equation}
    \partial_\tau q_1 = D(\bar\rho) \partial_x^2 q_1
    - \sigma(\bar\rho) \partial_x^2 p_1
    \:,
    \quad
    \partial_\tau p_1 = -D(\bar\rho) \partial_x^2 p_1
    \:,
\end{equation}
with
\begin{equation}
    p_1(x,t_0) = \frac{1}{\bar\rho} \Theta(x)
    \:,
    \quad
    q_1(x,0) = \frac{\sigma_i(\bar\rho)}{2 D_i(\bar\rho)}
    \left( p_1(x,0) - \frac{1}{\bar\rho} \Theta(x) \right)
    \:.
\end{equation}
These equations can be solved explicitly to give
\begin{equation}
    p_1(x,t) = \frac{1}{2\bar\rho}
    \erfc \left( - \frac{x}{\sqrt{4 D(\rho) (t_0 - t)}} \right)
    \:,
\end{equation}
\begin{multline}
    q_1(x,t) = \frac{\sigma_i(\bar\rho)}{4\bar\rho D_i(\bar\rho)} \left[
    \erfc \left( \frac{x}{\sqrt{4 D(\rho) t}} \right)
    - \erfc \left( \frac{x}{\sqrt{4 D(\rho) (t+t_0)}} \right)
    \right]
    \\
    + \frac{\sigma(\bar\rho)}{4\bar\rho D(\bar\rho)} \left[
    \erfc \left( \frac{x}{\sqrt{4 D(\rho) (t+t_0)}} \right)
    - \erfc \left( \frac{x}{\sqrt{4 D(\rho) (t_0-t)}} \right)
    \right]
    \:.
\end{multline}
Inserting these expressions into the definition of $Y[\rho]$~\eqref{eq:defXtfromRhoSM}, we get
\begin{equation}
    Y_1
    =
    \frac{1}{\bar\rho} \int_0^\infty [q_1(x,t_0) - q_1(x,0)] \dd x
    = \sqrt{t_0} \frac{\sqrt{2} D_i(\bar\rho) \sigma(\bar\rho)
    + (2 - \sqrt{2}) D(\rho) \sigma_i(\bar\rho)
    }{\bar\rho^2 D_i(\bar\rho) \sqrt{4 \pi D(\rho)}}
    \:.
\end{equation}

From these results, we deduce
\begin{equation}
    \label{eq:q1T1SM}
    q_1(x,t_0) = \frac{\sigma_i(\bar\rho)}{4\bar\rho D_i(\bar\rho)} \left[
    \erfc \left( \frac{x}{\sqrt{4 D(\rho) t_0}} \right)
    - \erfc \left( \frac{x}{\sqrt{8 D(\rho) t_0}} \right)
    \right]
    \\
    + \frac{\sigma(\bar\rho)}{4\bar\rho D(\bar\rho)} \left[
    \erfc \left( \frac{x}{\sqrt{8 D(\rho) t_0}} \right)
    - 2 \Theta(-x)
    \right]
    \:,
\end{equation}
which is a function of $x/\sqrt{t_0}$ only, thus
\begin{equation}
    \label{eq:Phi1SM}
    \Phi_1(z) \equiv q_1(z \sqrt{t_0},t_0)
\end{equation}
does not depend on $t_0$, as showed above.

Furthermore, having determined the profile $\Phi_1$, one can deduce the second cumulant of $X_t$ using relation~(8) from the main text, obtained in~\cite{Grabsch:2024bSM}, which gives
\begin{equation}
    \frac{1}{\sqrt{t}} \moy{X_t}_c^2 \frac{\lambda^2}{2}
    + O(\lambda^3)
    \underset{t \to \infty}{\simeq}
    -2 \frac{P'(\bar\rho) \Phi_1'(0^+) \lambda}{k_B T \bar\rho}
    D(\bar\rho) (\Phi_1(0^+) - \Phi_1(0^-))\lambda
    + O(\lambda^3)
    \:.
\end{equation}
Using the explicit expression of $\Phi_1$~(\ref{eq:q1T1SM},\ref{eq:Phi1SM}) together with $P'(\rho) = 2 \rho D(\rho)/\sigma(\rho)$ yields
\begin{equation}
    \label{eq:Xt2SM}
    \frac{1}{\sqrt{t}} \moy{X_t}_c^2
    \underset{t \to \infty}{\simeq} \frac{\sqrt{2} D_i(\bar\rho) \sigma(\bar\rho)
    + (2 - \sqrt{2}) D(\rho) \sigma_i(\bar\rho)
    }{\bar\rho^2 D_i(\bar\rho) \sqrt{4 \pi D(\rho)}}
    \:,
\end{equation}
which coincides with Eq.~(15) in the main text.

Note that this results can actually be obtained within MFT, since
\begin{equation}
    \dt{}{\lambda} \ln \moy{\e^{\lambda X_t}}
    \underset{\Lambda \to \infty}{\simeq}
    -\Nsc \dt{}{\lambda} (S[q,p] - \lambda Y[q] + F[q(x,0)])
    = \Nsc Y[q],
\end{equation}
since $(q,p)$ minimises $S - \lambda Y +F$. Therefore,
\begin{equation}
    \dt{}{\lambda} \ln \moy{\e^{\lambda X_t}}
    \equiv \lambda \moy{X_t^2}_c + O(\lambda^2)
    \simeq \Nsc Y_1 \lambda  + O(\lambda^2)
    = \underbrace{\Nsc \sqrt{t_0}}_{=\sqrt{t}}
    \frac{\sqrt{2} D_i(\bar\rho) \sigma(\bar\rho)
    + (2 - \sqrt{2}) D(\rho) \sigma_i(\bar\rho)
    }{\bar\rho^2 D_i(\bar\rho) \sqrt{4 \pi D(\rho)}} \lambda
     + O(\lambda^2)
    \:,
\end{equation}
which indeed coincides with~\eqref{eq:Xt2SM}.

\subsection{The case of a quenched initial condition}

For the sake of consistency, and to emphasise the difference with our computation, we reproduce here the result for the first-order correlation profile $\Phi_1$ in the case of a usual ``quenched'' initial condition, for which the initial density is not random, but fixed to the mean density~\cite{Krapivsky:2014SM,Krapivsky:2015aSM}
\begin{equation}
    \rho(x,0) = \bar\rho
    \:.
\end{equation}
The MFT approach works in the same way, and yields the same evolution equations~\eqref{eq:MFTbulkSM} and final condition~\eqref{eq:MFTfinalSM}, but with the initial condition~\eqref{eq:MFTinitSM} now replaced by
\begin{equation}
    q(x,0) = \bar\rho
    \:.
\end{equation}
Performing the same expansion in $\lambda$ as in~\ref{sec:MFTorder1}, the solution at first order reads~\cite{Krapivsky:2014SM,Krapivsky:2015aSM}
\begin{equation}
    q_1(x,t) =
     \frac{\sigma(\bar\rho)}{4\bar\rho D(\bar\rho)} \left[
    \erfc \left( \frac{x}{\sqrt{4 D(\rho) (t+t_0)}} \right)
    - \erfc \left( \frac{x}{\sqrt{4 D(\rho) (t_0-t)}} \right)
    \right]
    \:.
\end{equation}
The first correlation profile is thus
\begin{equation}
    \Phi_1(z) = q_1(z\sqrt{t_0},t_0) =
    \frac{\sigma(\bar\rho)}{4\bar\rho D(\bar\rho)} \left[
    \erfc \left( \frac{z}{\sqrt{8 D(\rho)}} \right)
    - 2 \Theta(-z)
    \right]
    \:.
\end{equation}
This function involves only the term $\propto \erfc ( \frac{z}{\sqrt{8 D(\bar\rho)}} )$, as claimed in the main text.

\subsection{Higher-order correlation profiles and cumulants}

Solving the MFT equations~(\ref{eq:MFTbulkSM}-\ref{eq:MFTinitSM}) for higher orders in $\lambda$ is more tricky, in part due to the appearance of $q(Y[q],t_0)$ in Eqs.~(\ref{eq:MFTfinalSM},\ref{eq:MFTinitSM}), which is ill-defined because $q(x,t_0)$ is discontinuous at $x = Y[q]$, as can be seen in the solution $q_1$ at first order~\eqref{eq:q1T1SM}. For the solution at first order in $\lambda$ this was not a problem, since only the order zero of $q(x,t_0) = \bar\rho + O(\lambda)$ was involved. For higher orders, this requires a specific treatment~\cite{Krapivsky:2015aSM}.

After the first works on tracer diffusion within MFT~\cite{Krapivsky:2014SM,Krapivsky:2015aSM}, other approaches have been implemented to overcome this difficulty, see for instance the discussion in Ref.~\cite{Grabsch:2024bSM}. In particular, this has permitted to compute the correlation profiles at orders $2$ and $3$ in $\lambda$, for arbitrary $D(\rho)$ and $\sigma(\rho)$, in the case of an annealed initial condition $T_i = T$~\cite{Grabsch:2024bSM}. We reproduce here these results for the generating function of the correlation profiles
\begin{equation}
    \Phi(z;\lambda) = \bar\rho + \lambda \Phi_1(z) + \frac{\lambda^2}{2} \Phi_2(z)
    + \frac{\lambda^3}{3!} \Phi_3(z) + \cdots
    \:,
\end{equation}
where
\begin{equation}
    \label{eq:Phi1bSM}
    \Phi_1(z) = \frac{\sigma(\rb)}{4 \rb D(\rb)}
    \erfc \left(y  \equiv \frac{z}{\sqrt{4 D(\rb)}} \right)
    \:,
\end{equation}
\begin{multline}
    \label{eq:Phi2SM}
   \Phi_2(z) =
   \frac{\sigma (\rb) \sigma '(\rb)}{8 \rb ^2 D(\rb )^2}
   \erfc(y)
   -\frac{\sigma (\rb)^2}{2 \pi  \rb ^3 D(\rb )^2}e^{-y^2}
   \\
   + \frac{\sigma (\rb )^2 D'(\rb)}{16 \pi  \rb^2 D(\rb)^3}
   \left[
   \pi  \erf(y) \erfc(y)+2 \sqrt{\pi } e^{-y^2} y \erfc(y)-2 (e^{-2 y^2}+e^{-y^2})
   \right]
   \:,
\end{multline}
\begin{multline}
    \label{eq:Phi3SM}
    \Phi_3(z) =
   \frac{3\sigma(\rb)^3}{8\pi ^{3/2} \rb^5 D(\rb)^3}
   \left(\sqrt{\pi} \text{erfc}(y)+2 e^{-y^2} y\right)
   +\frac{\sigma (\rb) \sigma'(\rb)^2}{16 \rb^3 D(\rb)^3}\text{erfc}(y)
   -\frac{\sigma (\rb)^2 \sigma '(\rb)}{16 \rb^4 D(\rb)^3}
   \left(\text{erfc}(y)+\frac{6 e^{-y^2}}{\pi }\right)
   \\
   +\frac{\sigma (\rb)^2\sigma ''(\rb)}{64 \rb^3 D(\rb)^3}
   \left(3 \text{erfc}\left(\frac{y}{\sqrt{2}}\right)^2-2 \text{erfc}(y)\right)
   -\frac{\sigma (\rb)^2 D'(\rb) \sigma'(\rb)}{32 \rb^3 D(\rb)^4}
   \left(-\frac{6 e^{-y^2} y \text{erfc}(y)}{\sqrt{\pi}}+3\text{erfc}(y)^2-\text{erfc}(y)+\frac{6 e^{-2 y^2}}{\pi }\right)
    \\
    +\frac{\sigma(\rb)^3 D'(\rb)}{32 \pi^{3/2} \rb^4 D(\rb)^4}
    \left(-12 \sqrt{\pi } e^{-y^2} y^2 \text{erfc}(y)+18 \sqrt{\pi } e^{-y^2}
   \text{erfc}(y)-\pi ^{3/2} \text{erfc}(y)+12 \sqrt{\pi } \text{erfc}(y)+12 e^{-2 y^2}
   y+12 e^{-y^2} y-12 \sqrt{\pi } e^{-y^2}\right)
   \\
   +\frac{3\sigma (\rb)^3 D'(\rb)^2}{128 \rb^3 D(\rb)^5}
   \Bigg(-\frac{4 e^{-2 y^2} y^2 \text{erfc}(y)}{\pi }-
   \frac{4 e^{-y^2} y^2 \text{erfc}(y)}{\pi }
   +\frac{4 e^{-y^2} y \text{erfc}(y)}{\sqrt{\pi }}
   +\frac{6 e^{-y^2} \text{erfc}(y)}{\pi}
   +\frac{2 e^{-y^2} y^3 \text{erfc}(y)^2}{\sqrt{\pi }}
   -2 \text{erfc}(y)^2
   \\
   +\frac{4\text{erfc}(y)}{\pi }+\frac{2 e^{-3 y^2} y}{\pi ^{3/2}}+\frac{4 e^{-2 y^2} y}{\pi
   ^{3/2}}+\frac{2 e^{-y^2} y}{\pi ^{3/2}}-\frac{4 e^{-2 y^2}}{\pi }
   \Bigg)
   \\
   +\frac{3\sigma (\rb)^3 \left(2 D(\rb) D''(\rb)-3 D'(\rb)^2\right)}{128 \rb ^3 D(\rb)^5}
   \left(-\frac{e^{-y^2}  y \text{erfc}(y)^2}{\sqrt{\pi }}-\frac{\sqrt{3} \text{erfc}(y)}{\pi}-\text{erfc}(y)+\text{erfc}\left(\frac{y}{\sqrt{2}}\right)^2
   +\frac{\sqrt{3}}{\pi }\text{erfc}\left(\sqrt{3} y\right)
   +\frac{2 e^{-y^2}}{3 \pi} +2 f(y)
   \right)
   \\
   -\frac{\sigma (\rb)^3 \left(D(\rb) D''(\rb)-3 D'(\rb)^2\right)}{64 \rb^3 D(\rb)^5}
   \left(-\frac{6 e^{-y^2} y \text{erfc}(y)^2}{\sqrt{\pi }} +\frac{6 e^{-2
   y^2} \text{erfc}(y)}{\pi }+\text{erfc}(y)^3-\frac{4 e^{-y^2}}{\pi }\right)
   \:,
\end{multline}
where the function $f$ is defined as
\begin{equation}
    \label{eq:FctFSM}
    f(y) = \int_0^1 \dd t \left[
       \frac{\e^{-\frac{y^2}{1-t^2}}}{\pi \sqrt{1-t^2}} \:
       \erf \left(
       \frac{t y}{\sqrt{(1-t^2)(1+2t)}}
       \right)
       - \frac{y(2+t)}{(1-t)(1+2t)^3/2 \pi^{3/2}}
       \e^{-\frac{(1+t)y^2}{(1-t)(1+2t)}}
    \right]
    \:.
\end{equation}
From these profiles, the higher-order cumulants can be obtained from Eq.~(8) in the main text, and are given by~\cite{Grabsch:2024bSM}
\begin{equation}
    \moy{X_t^3}_c = 0
    \:,
\end{equation}
\begin{multline}
    \label{eq:Kappa4SM}
    \frac{\moy{X_t^4}_c}{\sqrt{t}}
    \underset{t \to \infty}{\simeq}
    \frac{3 \sigma (\rb )^3 \left(\rb  D'(\rb )+D(\rb )\right)}
    {\pi ^{3/2} \rb^6 D(\rb )^{7/2}}
   -\frac{\sigma (\rb )\left(
   \sigma (\rb )\sigma'(\rb ) \left(\rb  D'(\rb )+4 D(\rb )\right)
   + 2 \sigma (\rb )^2 D'(\rb )-\rb  D(\rb ) \sigma'(\rb )^2
   \right)}
   {4 \sqrt{\pi} \rb^5 D(\rb )^{7/2}}
   \\
   +\frac{3 \sigma (\rb )^3  \left(D'(\rb )^2-D(\rb ) D''(\rb )\right)}
   {8 \sqrt{\pi } \rb^4 D(\rb)^{9/2}}
   +\frac{3 \sigma (\rb)^3 \left(2 D(\rb ) D''(\rb )-D'(\rb )^2\right)}
   {8 \pi ^{3/2} \rb ^4 D(\rb )^{9/2}}
   +\frac{\left(3 \sqrt{2}-4\right) \sigma (\rb )^2 \sigma ''(\rb )}
   {8 \sqrt{\pi } \rb ^4 D(\rb )^{5/2}}
   \\
   + \frac{3 \left(\sqrt{2} \pi -2 \sqrt{3}\right) \sigma (\rb)^3 \left(2 D(\rb ) D''(\rb )-3 D'(\rb )^2\right)}
   {16 \pi^{3/2} \rb ^4 D(\rb)^{9/2}}
    \:.
\end{multline}
These are the expressions used in the main text to compare to the numerical simulations.

\section{The transport coefficients}

The macroscopic description relies on the determination of the transport coefficients $D(\rho)$ and $\sigma(\rho)$. Although the expressions can be found in the literature~\cite{Lekkerkerker:1981SM,Cichocki:1991SM,Butta:1999SM,Felderhof:2009SM}, we briefly sketch here their derivation for interacting Brownian particles, before applying them to the Calogero potential and the WCA potential.

\subsection{Expression for interacting Brownian particles}

First, it is well-known that $D(\rho)$ and $\sigma(\rho)$ are related by the fluctuation-dissipation relation~\cite{Bertini:2015SM}
\begin{equation}
    \label{eq:RelDSigmaFSM}
    \frac{2 k_B T D(\rho)}{\sigma(\rho)} = f''(\rho)
    \:,
\end{equation}
with $f(\rho)$ the equilibrium free energy density, for a system at density $\rho$, and $f''$ its second derivative with respect to the density $\rho$. Since the ratio $D/\sigma$ is known, we only need to determine one of the coefficients to obtain the second. In many situations, it is simpler to determine $D(\rho)$ and deduce $\sigma(\rho)$, see for instance~\cite{Arita:2017SM,Arita:2018SM}. Here, for interacting Brownian particles, it is simpler to first determine $\sigma(\rho)$. Indeed, $\sigma$ can be seen as the linear response of the mean current $\moy{j}$ to the application of an external force $F_0$ on all the particles,
\begin{equation}
\label{eq:defSigmaSM}
    \moy{j} = \frac{\sigma(\rho)}{2 k_B T} F_0
    \:.
\end{equation}
The addition of the force $F_0$ changes the Langevin equations~\eqref{eq:EqBrownianPartSM} into
\begin{equation}
    \label{eq:EqBrownianPartForceSM}
    \dt{x_i}{t} = - \mu_0 \sum_{j \neq i} V'(x_i-x_j)
    + \mu_0 F_0
    + \sqrt{2 D_0} \: \eta_i
    \:.
\end{equation}
Since $F_0$ is constant, it can be absorbed by a change of reference frame,
\begin{equation}
    y_i(t) = x_i(t) + v t
    \:,
    \quad
    v = \mu_0 F_0
    \:,
\end{equation}
where the $y_i$'s obey the original equations~\eqref{eq:EqBrownianPartSM}. Since the system described by~\eqref{eq:EqBrownianPartSM} is symmetric under $x_i \to - x_i$ (for symmetric $V$), there is no mean displacement of the particles in the moving frame $\moy{y_i(t)} = \moy{y_i(0)}$, and thus the mean current in the fixed reference frame is
\begin{equation}
    \moy{j} = \rho v = \rho \mu_0 F_0
    \:.
\end{equation}
Combining this expression with the definition of $\sigma$~\eqref{eq:defSigmaSM} yields
\begin{equation}
    \label{eq:SigmaSM}
    \sigma(\rho) = 2 \mu_0 k_B T  \rho
    \:,
\end{equation}
corresponding to Eq.~(4), left, in the main text.

The diffusion coefficient can be deduced from~\eqref{eq:RelDSigmaFSM},
\begin{equation}
    \label{eq:D0SM}
    D(\rho) = \mu_0 \rho f''(\rho)
    \:.
\end{equation}
Using standard relations of thermodynamics~\cite{Hill:1986SM}, we have for a system of volume $V$ with $N = \rho V$ particles,
\begin{equation}
    \label{eq:ThermoSM}
 P = - \left( \frac{\partial F}{\partial V} \right)_{T,N}
 = -  \left( \frac{\partial [V f(N/V)]}{\partial V} \right)_{T,N}
 = \rho f'(\rho) - f(\rho)
 \:,
\end{equation}
with $F = V f$ the Helmholtz free energy and $P$ the pressure. We thus have that
\begin{equation}
    \partial_\rho P = \rho f''(\rho)
    \:.
\end{equation}
Identifying with~\eqref{eq:D0SM}, we obtain
\begin{equation}
    \label{eq:DSM}
    D(\rho) = \mu_0 \: \partial_\rho P(\rho)
    \:,
\end{equation}
which is indeed Eq.~(4), right, in the main text.

\bigskip

Note that the analysis performed above can be straightforwardly extended in arbitrary dimension, since a system of pairwise interacting Brownian particles submitted to an external force $\boldsymbol{F}_0$ in $d$ dimensions is described by
\begin{equation}
    \dt{\boldsymbol{x}_i}{t} =
    - \mu_0 \sum_{j\neq i} \boldsymbol{\nabla} V(\boldsymbol{x}_i - \boldsymbol{x}_j)
    + \mu_0 \boldsymbol{F}_0
    + \sqrt{2 D_0} \boldsymbol{\eta}_i
    \:,
\end{equation}
with $\boldsymbol{\eta}_i$ a $d$-dimensional vector of white noises. The force can be absorbed by a change of frame with velocity $\boldsymbol{v} = \mu_0 \boldsymbol{F}_0$, indicating that the mean current is $\moy{\boldsymbol{j}} = \rho \mu_0 \boldsymbol{F}_0$, and thus $\sigma(\rho)$ is still given by~\eqref{eq:SigmaSM}. Furthermore, since both the fluctuation-dissipation relation~\eqref{eq:RelDSigmaFSM} and the thermodynamic identities~\eqref{eq:ThermoSM} hold in any dimension, so does the expression of the diffusion coefficient $D(\rho)$~\eqref{eq:DSM}.

\bigskip

The expression of the diffusion coefficient requires the knowledge of the equation of state of the system, which is known exactly for specific models, or can be approximated with classical methods for an arbitrary potential $V$.

\subsection{A comment on the thermodynamic quantities}
\label{sec:CommentThermoSM}

In standard statistical mechanics~\cite{Hill:1986SM}, the free energy $F$ is defined from the canonical partition function $Z_N$ as
\begin{equation}
    F \underset{N \to \infty}{\simeq} - k_B T \ln Z_N
    \:.
\end{equation}
For a system of $N$ particles of mass $m$ pairwise interacting via the potential $V$ in a system of length $L$, the partition function is expressed as
\begin{equation}
    \label{eq:PartFctSM}
    Z_N(\beta) = \frac{1}{h^N N!} \int_{-\infty}^\infty \dd p_1 \cdots \dd p_N
    \int_0^L \dd x_1 \cdots \dd x_N \: \e^{- \sum_{i=1}^N \frac{\beta p_i^2}{2m}
    - \frac{\beta}{2} \sum_{i \neq j} V(x_i - x_j)}
    \:,
\end{equation}
with $h$ the Planck constant. However, here the dynamics of the particles is described by~\eqref{eq:EqBrownianPartSM}, and their mass is irrelevant (and so is $h$). Thus, what sense do we give to the pressure in~\eqref{eq:DSM}?

\bigskip

It turns out that the pressure in~\eqref{eq:DSM} can be understood as the pressure in a classical ensemble of particles described by the standard partition function~\eqref{eq:PartFctSM}, because the equation of state obtained from~\eqref{eq:PartFctSM} does not involve $m$ or $h$. Since this will be useful in the following, we briefly re-derive this important property here.

\bigskip

First, the integral over the momenta in~\eqref{eq:PartFctSM} can be performed explicitly, to yield
\begin{equation}
    \label{eq:ZNSM}
    Z_N(\beta) = \frac{1}{N! \ell_0^N}
    \int_0^L \dd x_1 \cdots \dd x_N \:
    \e^{- \frac{\beta}{2} \sum_{i \neq j} V(x_i - x_j)}
    \:,
    \quad \text{with} \quad
    \ell_0 = \sqrt{ \frac{ \beta h^2}{2m \pi} }
    \:.
\end{equation}
The grand canonical partition function thus takes the form
\begin{equation}
    \label{eq:GrandPartitionSM}
    \mathcal{Z}(\beta, \varphi) = \sum_{N=0}^\infty \varphi^N Z_N
    = \sum_{N=0}^\infty \frac{1}{N!} \left(\frac{\varphi}{\ell_0} \right)^N
    \int_0^L \dd x_1 \cdots \dd x_N \:
    \e^{- \frac{\beta}{2} \sum_{i \neq j} V(x_i - x_j)}
    \:,
\end{equation}
where $\varphi = \e^{\beta \mu}$ is the fugacity, and $\mu$ the chemical potential.
We introduce the grand potential density $\phi_G$ as
\begin{equation}
    \phi_G L \underset{L \to \infty}{\simeq} - k_B T \ln \mathcal{Z}(\beta, \varphi)
    \:.
\end{equation}
From~\eqref{eq:GrandPartitionSM}, the grand potential is a function of $\beta$ and $\varphi/\ell_0$,
\begin{equation}
    \label{eq:ScalingGrandPotSM}
    \phi_G(\beta,\varphi) \equiv \tilde{\phi}_G(\beta, \varphi/\ell_0 \equiv \tilde{\varphi})
    \:.
\end{equation}
The grand potential is directly related to the pressure via $\phi_G = - P$. However, $\phi_G$, and thus $P$, are for the moment expressed as functions of the chemical potential $\mu$ (via the fugacity $\varphi$). We need to express them in terms of the density $\rho$. To do so, we go back to the canonical ensemble by using that
\begin{equation}
    Z_N(\beta) = \oint \frac{\dd \varphi}{2\I \pi \varphi^{N+1}} \mathcal{Z}(\beta,\varphi)
    \simeq \int \dd \mu \: \e^{ -\beta L( \phi_G(\beta,\e^{\beta \mu}) + \mu \rho)}
    \:,
\end{equation}
where we have denoted $\rho = N/L$. Performing a saddle-point estimation of the last integral for large $L$, we recover that the free energy and the grand potential are related by a Legendre transform,
\begin{equation}
    f(\rho) \equiv F/L = \phi_G(\beta, \e^{\beta \mu^\star}) + \rho \mu^\star
    \:,
    \quad
    \partial_\mu \phi_G(\beta, \e^{\beta \mu} ) \Bigg|_{\mu^\star} = - \rho
    \:.
\end{equation}
Finally, this gives the expression of the pressure in terms of the density,
\begin{equation}
    P(\rho) = - \phi_G \left(\beta, \e^{\beta \mu^\star(\beta,\rho)} \right)
    \:.
\end{equation}
Using that the grand potential depends on the length $\ell_0$ through~\eqref{eq:ScalingGrandPotSM}, this becomes
\begin{equation}
    \label{eq:PressureSM}
    P(\rho) = - \tilde\phi_G \left(\beta, \e^{\beta \mu^\star(\beta,\rho)}/\ell_0 \right)
    \:,
    \quad \text{with} \quad
    \beta \frac{\e^{\beta \mu^\star}}{\ell_0} \partial_{\tilde{\varphi}} \tilde{\phi}_G \Big|_{\tilde{\varphi} = \e^{\beta \mu_\star}/\ell_0} = -\rho
    \:.
\end{equation}
The second equation shows that $\e^{\beta\mu^\star}/\ell_0$ is a function of the density $\rho$ only, and since the pressure $P$ depends on $\mu_\star$ and $\ell_0$ only through the combination $\e^{\beta\mu^\star}/\ell_0$, it is only a function of $\rho$, and does not depend on $\ell_0$. Therefore, even if the free energy or the grand potential depend explicilty on the masses $m$ of the particles or on Planck's constant $h$, the pressure $P(\rho)$ expressed as a function of the density does not depend on these quantities. $P(\rho)$ depends only on the parameters involved in the overdamped Langevin equations~\eqref{eq:EqBrownianPartSM}, and so does the diffusion coefficient $D(\rho)$~\eqref{eq:DSM}, as it should.

\subsection{The Calogero potential}

We consider here a one-dimensional system of Brownian particles interacting with the Calogero potential
\begin{equation}
    \label{eq:CalogeroSM}
    V(x) = \frac{a}{x^2}
    \:.
\end{equation}

For a classical system of $N$ particles of mass $m$ in an interval of length $L$, the partition function
\begin{equation}
    Z_N = \frac{1}{N! h^N} \int_{0}^L \dd x_1 \cdots \dd x_N
    \: \e^{- \frac{\beta}{2} \sum_{i \neq j} V(x_i - x_j)}
    \int \dd p_1 \cdots \dd p_N \e^{-\sum_{i=1}^N\frac{\beta p_i^2}{2m}}
    \:,
\end{equation}
has been computed in the thermodynamic limit~\cite{Choquard:2000SM}. More precisely, the grand partition function
\begin{equation}
    \mathcal{Z} = \sum_{N=0}^\infty \varphi^N Z_N
\end{equation}
takes the form~\cite{Choquard:2000SM} (see also the review~\cite{Lewin:2022SM})
\begin{equation}
    \label{eq:GPCaloSM}
    \frac{k_B T}{L} \ln \mathcal{Z}
    \underset{L \to \infty}{\simeq}
    \frac{1}{\pi \beta\sqrt{2 \beta a}} \int_{-\infty}^\infty
    W \left( 2 \pi \frac{\sqrt{a m}}{h} \varphi \e^{-k^2} \right) \dd k
    = \frac{1}{\pi \beta \sqrt{2 \beta a}} \int_{-\infty}^\infty
    W \left(  \sqrt{2 \pi \beta a } \frac{\varphi}{\ell_0} \e^{-k^2} \right) \dd k
    \equiv -\phi_G(\beta,\varphi)
    \:,
\end{equation}
where $W$ is the Lambert-W function which satisfies $W(x) \e^{W(x)} = x$, and $\ell_0 = \sqrt{\beta h^2 / (2 \pi m)}$ is the length scale introduced in~\eqref{eq:ZNSM}. As discussed in Section~\ref{sec:CommentThermoSM}, the grand potential density~\eqref{eq:GPCaloSM} depends on the length $\ell_0$, but the pressure obtained from it does not. Indeed from~\eqref{eq:PressureSM}, we have that
\begin{equation}
    \label{eq:PCaloSM}
    P(\rho) = \frac{1}{\pi \beta \sqrt{2 \beta a}} \int_{-\infty}^\infty
    W \left(  \sqrt{2 \pi \beta a } \: \tilde\varphi^\star(\rho) \e^{-k^2} \right) \dd k
    \:,
\end{equation}
where $\tilde{\varphi}^\star(\rho)$ is solution of
\begin{equation}
    \label{eq:PparamCaloSM}
    \sqrt{\pi} \tilde\varphi(\rho) \int_{-\infty}^\infty
    W'\left(  \sqrt{2 \pi \beta a } \: \tilde\varphi(\rho) \e^{-k^2} \right) \e^{-k^2} \dd k
    = \rho
    \:,
\end{equation}
and $W'(x) = W(x)/(x(1+W(x)) )$ is the derivative of the Lambert-W function. These expressions no longer depend on $\ell_0$, as showed above, and involve only the parameters $\beta$ and $a$ which characterise the microscopic model. The diffusion coefficient obtained from~(\ref{eq:PCaloSM},\ref{eq:PparamCaloSM}), together with~\eqref{eq:DSM}, is shown in Fig.~\ref{fig:DiffCoefsSim} below.

\bigskip

The expressions~(\ref{eq:PCaloSM},\ref{eq:PparamCaloSM}) determine the pressure $P(\rho)$ parametrically (via $\tilde\varphi^\star$). These expressions can be analysed in various limits to obtain asymptotic expressions for $P(\rho)$, and thus $D(\rho)$. For instance, for $\beta \to \infty$ or $\rho \to \infty$, denoting $\varphi^\star = \e^{\beta \mu^\star}/\ell_0$, Eqs.~(\ref{eq:PCaloSM},\ref{eq:PparamCaloSM}) become
\begin{equation}
    P(\rho) \simeq \frac{4 (\mu^\star)^{3/2}}{3 \pi \sqrt{2 a}}
    \:,
    \quad
    \frac{\sqrt{2\mu^\star}}{ \pi \sqrt{a}} = \rho
    \:,
\end{equation}
hence
\begin{equation}
    P(\rho) \simeq \frac{\pi^2  \rho^3 a}{3}
    \:,
    \quad \text{for} \quad
    T \to 0
    \quad \text{or} \quad
    \rho \to \infty
    \:.
\end{equation}
The diffusion coefficient of the Calogero gas in this limit is obtained from~\eqref{eq:DSM},
\begin{equation}
    \label{eq:DCaloHighDens}
    D(\rho) \simeq \mu_0 \pi^2  \rho^2 a
    \:,
    \quad \text{for} \quad
    T \to 0
    \quad \text{or} \quad
    \rho \to \infty
    \:.
\end{equation}
Inserting this expression into Eq.~(15) of the main text for $T_i = T$, we recover the low temperature behaviour of $\moy{X_t^2}$ recently found in~\cite{Touzo:2024SM,Touzo:2024aSM}. Note that, in this regime, we also recover the same result for the diffusion coefficient as in~\cite{Dandekar:2023SM}. Here, we also have access to the other asymptotic regime of $\beta \to 0$ or $\rho \to 0$. In this case, Eqs.~(\ref{eq:PCaloSM},\ref{eq:PparamCaloSM}) can be expanded in powers of $\tilde{\varphi}^\star$ to give
\begin{equation}
    P(\rho) = \frac{\tilde{\varphi}^\star}{\beta} - \sqrt{\frac{a \pi}{\beta}} (\tilde{\varphi}^\star)^2
    + \sqrt{3} \pi a (\tilde{\varphi}^\star)^3
    + \cdots
    \:,
    \quad
    \rho = \tilde{\varphi}^\star - 2 \sqrt{\pi \beta a} (\tilde{\varphi}^\star)^2
    + 3 \sqrt{3} \pi \beta a (\tilde{\varphi}^\star)^3 + \cdots
    \:.
\end{equation}
Inverting the second series and inserting it into the first gives
\begin{equation}
    \label{eq:PCaloLowDensSM}
    P(\rho) = \frac{\rho}{\beta} + \sqrt{\frac{a \pi}{\beta}} \rho^2
    + 2 (2-\sqrt{3}) a \pi \rho^3 + O(\rho^4)
    \quad \text{for} \quad
    T \to \infty
    \quad \text{or} \quad
    \rho \to 0
    \:.
\end{equation}
Inserting this expression into~\eqref{eq:DSM}, we obtain
\begin{equation}
    D(\rho) = \mu_0 k_B T
    + 2 \mu_0 \sqrt{\pi k_B T a} \rho
    + 6 \pi (2-\sqrt{3})\mu_0  a \rho^2
    + O(\rho^3)
    \:,
    \quad \text{for} \quad
    T \to \infty
    \quad \text{or} \quad
    \rho \to 0
    \:.
\end{equation}

\subsection{The virial expansion}

For an arbitrary choice of the interaction potential $V(x)$, no expression for the pressure $P(\rho)$ is available. Nevertheless, many approximation schemes have been developed to obtain an approximate equation of state, such as the virial expansion~\cite{Hill:1986SM}. It consists in expanding the grand canonical partition function~\eqref{eq:GrandPartitionSM} in ``clusters'' which involve first two-body interactions, then three-body, etc. It yields an expression for the pressure in powers of the density as
\begin{equation}
    \label{eq:PressureVirialSM}
    \frac{P(\rho)}{k_B T} = \rho + B_2(T) \rho^2 + B_3(T) \rho^3 + O(\rho^4)
    \:,
\end{equation}
where $B_n(T)$ is the $n^{\mathrm{th}}$ virial coefficient. The first coefficients have a simple form~\cite{Hill:1986SM},
\begin{equation}
    \label{eq:VirialB2SM}
    B_2(T) = - \frac{1}{2V} \int_V \dd^d \boldsymbol{x}_1 \int_V \dd^d \boldsymbol{x}_2
    \left( \e^{-\beta V(\boldsymbol{x}_1 - \boldsymbol{x}_2)} - 1\right)
    \underset{V \to \infty}{\simeq}
    - \frac{1}{2} \int \dd^d \boldsymbol{x}_1
    \left( \e^{-\beta V(\boldsymbol{x}_1)} - 1\right)
    \:,
\end{equation}
\begin{align}
    B_3(T)
    &= - \frac{1}{3V} \int_V \dd^d \boldsymbol{x}_1
    \int_V \dd^d \boldsymbol{x}_2
    \int_V \dd^d \boldsymbol{x}_3
    f_M(\boldsymbol{x}_1  - \boldsymbol{x}_2)
    f_M(\boldsymbol{x}_1  - \boldsymbol{x}_3)
    f_M(\boldsymbol{x}_2  - \boldsymbol{x}_3)
    \nonumber
    \\
    &\underset{V \to \infty}{\simeq}
    - \frac{1}{3} \int \dd^d \boldsymbol{r}_1
    \int \dd^d \boldsymbol{r}_2
    f_M(\boldsymbol{r}_1)
    f_M(\boldsymbol{r}_2)
    f_M(\boldsymbol{r}_2  - \boldsymbol{r}_1)
    \label{eq:VirialB3SM}
    \:,
\end{align}
where we introduced the Mayer function
\begin{equation}
    \label{def:MayerSM}
    f_M(\boldsymbol{x}) = \e^{-\beta V(\boldsymbol{x})} - 1
    \:.
\end{equation}
Since we consider potentials that are rotationally invariant, we can reduce the number of integrals to perform by going to spherical coordinates. Indeed, the coefficient $B_2$~\eqref{eq:VirialB2SM} becomes
\begin{equation}
    B_2(T) =- \frac{\pi^{d/2}}{\Gamma(\frac{d}{2})}
    \int_0^\infty r^{d-1} \left( \e^{-\beta V(r)} - 1\right) \dd r
    \:.
    \label{appeq:b2}
\end{equation}
In two dimensions, the third virial coefficient can be computed as
\begin{equation}
    B_3(T) = - \frac{2\pi}{3} \int_0^\infty \dd r_1
    \int_0^\infty \dd r_2
    \int_0^{2\pi} \dd \theta \:
    r_1 f_M(r_1)
    r_2 f_M(r_2)
    f_M \left(\sqrt{r_1^2 + r_2^2 - 2 r_1 r_2 \cos \theta} \right)
    \:.
    \label{appeq:b3}
\end{equation}

We will make use of the expansion~\eqref{eq:PressureVirialSM} to compare our results to numerical simulations, in particular for the WCA potential. But first, we discuss a few specific cases.

\subsubsection{The Calogero potential in one dimension}

Let us reconsider here the Calogero potential~\eqref{eq:CalogeroSM}. The second virial coefficient takes the form
\begin{equation}
    B_2(T) = \int_0^\infty (1 - \e^{- \beta a/x^{2}} ) \dd x
    = \sqrt{\beta a} \int_0^\infty (1 - \e^{- 1/u^{2}} ) \dd u
    = \sqrt{\pi \beta a}
    \:,
\end{equation}
which matches with the expansion~\eqref{eq:PCaloLowDensSM}. For the third virial coefficient, we have
\begin{equation}
    B_3(T) = \frac{\beta a}{3} \int_{-\infty}^\infty \dd u \int_{-\infty}^\infty \dd v
    (1-\e^{-1/u^2}) (1-\e^{-1/v^2}) (1-\e^{-1/(u-v)^2})
    \:.
\end{equation}
We have checked numerically that the integral yields the numerical factor
\begin{equation}
    B_3(T) = \beta a 2 \pi (2-\sqrt{3})
    \:,
\end{equation}
again in agreement with~\eqref{eq:PCaloLowDensSM}, as it should.

\subsubsection{The Riesz potential in one dimension}

The Calogero potential is part of a larger class of potentials, called the Riesz potentials, which have been the object of numerous works, see for instance the review~\cite{Lewin:2022SM}. These are potentials of the form
\begin{equation}
\label{eq:RieszSM}
    V(x) = \frac{a}{\abs{x}^s}
    \:.
\end{equation}
The second virial coefficient can be computed exactly,
\begin{equation}
    B_2(T) = \int_0^\infty (1 - \e^{- \beta a/x^s}) \dd x
    = (\beta a)^{1/s} \int_0^\infty (1 - \e^{- 1/u^s}) \dd u
    =  (\beta a)^{1/s} \Gamma \left( 1 - \frac{1}{s} \right)
    \:.
\end{equation}
Note that the integral converges only for $s>1$. If $s<1$, the divergence of $B_2$ indicates that the system no longer obeys the standard thermodynamics, in particular the energy is no longer extensive due to the long-range interactions. In this case, the system of particles described by the Langevin equations~\eqref{eq:EqBrownianPartSM} is no longer diffusive~\cite{Dandekar:2023SM} and the stochastic equation~\eqref{eq:flucHydroSM} no longer applies. We thus focus on the case $s>1$, for which the diffusion coefficient~\eqref{eq:DSM} reads
\begin{equation}
    \label{eq:RieszDiff}
    D(\rho) = \mu_0 k_B T \left[
    1 + 2 \Gamma \left( 1 - \frac{1}{s} \right) (\beta a)^{1/s}  \rho
    + O(\rho^2)
    \right]
    \:.
\end{equation}

\subsubsection{The WCA potential in one dimension}

We consider the Weeks-Chandler-Andersen potential, defined by
\begin{equation}
    \label{eq:WCASM}
    V(x) = a \left\lbrace
    \begin{array}{cc}
    \displaystyle
    4 \left[ \left( \frac{\ell}{x} \right)^{12} - \left( \frac{\ell}{x} \right)^{6} \right] + 1
    & \text{for } \abs{x} < \ell \: 2^{1/6}
    \:,
    \\[0.3cm]
    0 & \text{for } \abs{x} > \ell \: 2^{1/6}
    \:.
    \end{array}
    \right.
\end{equation}
Because the potential is not scale invariant, unlike the Riesz potential~\eqref{eq:RieszSM}, we cannot obtain a simple expression for the virial coefficients~(\ref{eq:VirialB2SM},\ref{eq:VirialB3SM}) that holds for any values of $a$ and $\beta$. In the numerical simulations (see below), we have mostly used $a=1$, $\beta = 1$ and $\ell = 1$, for which~(\ref{eq:VirialB2SM},\ref{eq:VirialB3SM}) give
\begin{equation}
    \frac{P(\rho)}{k_B T} =
    \rho + 1.01561 \rho^2 + 1.02977 \rho^3 +  O(\rho^4)
    \:,
\end{equation}
from which we deduce the diffusion coefficient~\eqref{eq:DSM}
\begin{equation}
    D(\rho) = \mu_0 k_B T \left(
    1 + 2.03121 \rho + 3.08932 \rho^2 + O(\rho^3)
    \right)
    \:.
\end{equation}
This expansion is compared to numerical simulations in Fig.~\ref{fig:DiffCoefsSim} below. For these values of $T$, $\ell$ and $a$, the agreement with the second order virial expansion is good for $\rho \leq 0.2$.

\subsubsection{The WCA potential in two dimensions}
In numerical simulations of two-dimensional systems (see Sec.~\ref{sec:simul-2d} below), we have used the
WCA interaction potential~\eqref{eq:WCASM} with $a=1$ and $\ell=1$. To describe temperature quenches from $T_i=1$ to $T=0.5$, we need to evaluate the first few virial coefficients $B_n(T)$ for both these situations. These can be evaluated numerically via Eqs.~\eqref{appeq:b2} and~\eqref{appeq:b3} to give (within 5 digits precision)
\begin{equation}
    B_2(1) = 1.62285, \quad B_3(1) =2.05007, \quad B_2(0.5) = 1.69752, \quad B_3(0.5) = 2.24659.
\end{equation}

\subsubsection{The Gaussian potential in one dimension}

We consider a Gaussian interaction potential
\begin{equation}
    \label{eq:GaussPotSM}
    V(x) = a \: \e^{- \frac{x^2}{\ell^2}}
    \:.
\end{equation}
In this case, there is no hardcore repulsion and the particles can bypass each other. For $a=2$, $\ell = 1$ and $T=1$, the virial expansion~\eqref{eq:PressureVirialSM} gives
\begin{equation}
    \frac{P(\rho)}{k_B T}
    = \rho + 0.657563 \rho^2 + 0.707915 \rho^3 + O(\rho^4)
    \:,
\end{equation}
so that
\begin{equation}
    \label{eq:VirialGaussSM}
    D(\rho) = \mu_0 k_B T \left(
    1 + 1.97269 \rho + 2.12375 \rho^2 + O(\rho^3)
    \right)
    \:.
\end{equation}
Note that the linearised Dean-Kawasaki equation gives~\cite{Demery:2014SM,Illien:2025SM}
\begin{equation}
    \label{eq:LinDKGaussSM}
    D(\rho) = \mu_0 k_B T + \mu_0 \rho \int V(x) \dd x
    = \mu_0 k_B T + \sqrt{\pi} \mu_0 a \ell \rho
    \:.
\end{equation}
These two expressions are compared to numerical simulations in Fig.~\ref{fig:DiffCoefsSim}. The virial expansion~\eqref{eq:VirialGaussSM} is a good approximation for low densities $\rho \leq 0.4$, while the linearised Dean-Kawasaki result~\eqref{eq:LinDKGaussSM} gives good results at high density $\rho \geq 0.6$, which is the expected domain of validity of the linearisation~\cite{Demery:2014SM}.

\subsubsection{The Gaussian potential in two dimensions}
In numerical simulations of two-dimensional systems (see Sec.~\ref{sec:simul-2d} below), we have used the
Gaussian interaction potential~\eqref{eq:GaussPotSM} with $a=1$ and $\ell=1$. To describe temperature quenches from $T_i=1$ to $T=0.1$, we need to evaluate the first few virial coefficients $B_n(T)$ for both these situations. These can be evaluated numerically to give (within 5 digits precision)
\begin{equation}
    B_2(1) = 1.25130, \quad B_3(1) =0.485043, \quad B_2(0.1) = 4.52359, \quad B_3(0.1) = 13.9997.
\end{equation}

\subsection{Comparison with the results of Dandekar, Krapivsky, Mallick\texorpdfstring{~\cite{Dandekar:2023SM}}{}}

In Ref.~\cite{Dandekar:2023SM}, Dandekar, Krapivsky and Mallick study the system of interacting Brownian particles~\eqref{eq:EqBrownianPartSM} with the Riesz potential~\eqref{eq:RieszSM} relying on a Martin-Siggia-Rose approach to the Dean-Kawasaki equation (2) combined with a saddle-point analysis. As discussed in the main text, this saddle-point computation lacks a clear large parameter to be fully controlled, while the MFT naturally
features the large rescaling parameter $\Lambda \gg 1$ that allows to obtain the exact large-scale behavior.
Ref.~\cite{Dandekar:2023SM} mostly focuses on the long-range case $s<1$, which yields a subdiffusive behavior at large scales, which is not captured by the usual MFT formalism because it corresponds to a non-standard thermodynamics (the energy is no longer extensive). In Ref.~\cite{Dandekar:2023SM}, the authors also discuss the short-range case $s>1$, which
instead
is captured by the MFT formalism used in this work. In particular, Eq.~(76) of~\cite{Dandekar:2023SM} gives the diffusion coefficient (rewritten in our notations)
\begin{equation}
    \label{eq:DKMRiesz}
    D(\rho) = \mu_0 k_{\mathrm{B}} T
    + \mu_0 s(1+s) \zeta(s) a \: \rho^s
    \:.
\end{equation}

\noindent $\bullet$ At low density, this has to be compared to the exact linear low-density behavior obtained from the virial expansion~\eqref{eq:RieszDiff}. We see that Eq.~\eqref{eq:DKMRiesz} is unable to reproduce the exact low-density behavior of the diffusion coefficient.\\

\noindent $\bullet$ At arbitrary density,
in Fig.~\ref{fig:DiffCoefsSim} (left) below
we compare
Eq.~\eqref{eq:DKMRiesz} to
the exact expression for the pressure (and thus the diffusion coefficient)
that we
obtained in the specific case of the Calogero gas $s=2$~(\ref{eq:PCaloSM},\ref{eq:PparamCaloSM}). The numerical simulation unambiguously supports the result~(\ref{eq:PCaloSM},\ref{eq:PparamCaloSM}) and not Eq.~\eqref{eq:DKMRiesz}.
However, in the high-density (or low-temperature) regime, Eq.~\eqref{eq:DKMRiesz} matches the exact behavior of $D(\rho)$ given by~\eqref{eq:DCaloHighDens}.\\

To summarize, Eq.~\eqref{eq:DKMRiesz} describes the diffusion coefficient of the Riesz gas only in the high-density (or low-temperature) regime. This could be the sign of the emergence of a large parameter which renders the saddle-point calculation of~\cite{Dandekar:2023SM} well controlled  in this regime.

\section{Density-density correlations in arbitrary dimension}

The density-density correlations can be obtained directly in any dimension from the multidimensional version of the evolution equation~\eqref{eq:flucHydroSM}, which reads~\cite{Bertini:2015SM}
\begin{equation}
    \label{eq:flucHydroAnyDSM}
   \partial_\tau \rho =
   \boldsymbol{\nabla} \cdot
   \left[
    D(\rho) \boldsymbol{\nabla} \rho
    + \sqrt{\frac{\sigma(\rho)}{\Nsc^d} } \: \boldsymbol{\eta}
    \right]
    \:,
\end{equation}
where $\boldsymbol{\eta}$ is a vector of Gaussian white noises
\begin{equation}
    \moy{ \boldsymbol{\eta}_i(\boldsymbol{x},\tau) \boldsymbol{\eta}_j(\boldsymbol{y},\tau') }
    = \delta_{i,j} \delta( \boldsymbol{x}- \boldsymbol{y})
    \delta(\tau-\tau')
    \:.
\end{equation}
Since $\Nsc \gg 1$, we can look for a solution in powers of $\Nsc^{-1/2}$,
\begin{equation}
    \rho = \bar\rho + \frac{1}{\sqrt{\Nsc^d}} \rho_1 + O(\Nsc^{-d})
    \:.
\end{equation}
The evolution equation~\eqref{eq:flucHydroAnyDSM} thus becomes at leading order
\begin{equation}
    \partial_{\tau} \rho_1 = D(\bar\rho) \Delta \rho_1
    + \sqrt{\sigma(\bar\rho)}  \boldsymbol{\nabla} \cdot \boldsymbol{\eta}
    \:.
\end{equation}
This equation can be solved explicitly, and the solution reads
\begin{equation}
    \rho_1(\boldsymbol{x},\tau)
    = \sqrt{\sigma(\bar\rho)} \int_0^\tau \dd \tau' \int \dd^d \boldsymbol{y}
    K(\boldsymbol{x},\tau | \boldsymbol{y},\tau')
    \boldsymbol{\nabla} \cdot \boldsymbol{\eta}(\boldsymbol{y},\tau')
    + \int \dd^d \boldsymbol{z}
    K(\boldsymbol{x},\tau | \boldsymbol{z},0) \rho_1(\boldsymbol{z},0)
    \:,
\end{equation}
where the propagator is
\begin{equation}
    K(\boldsymbol{x},\tau | \boldsymbol{y},\tau')
    = \frac{\e^{- \frac{(\boldsymbol{x}- \boldsymbol{y})^2}{4 D(\bar\rho)(\tau-\tau')}}}{(4 \pi D(\bar\rho) (\tau-\tau') )^{d/2}}
    \:.
\end{equation}
Since the noise $\boldsymbol{\eta}$ is not correlated with the initial condition $\rho_1(\boldsymbol{y},0)$, we have that
\begin{equation}
    \label{eq:CorrelDensDens0SM}
    \moy{\rho_1(\boldsymbol{x},\tau) \rho_1(\boldsymbol{y},0)}_c
    = \int \dd^d \boldsymbol{y}
    K(\boldsymbol{x},\tau | \boldsymbol{y},0)
    \moy{ \rho_1(\boldsymbol{z},0) \rho_1(\boldsymbol{y},0)}_c
    \:.
\end{equation}
We thus only need to determine the correlations between two points for the initial data $\rho(\boldsymbol{x},0)$. We consider the situation in which the initial density is picked from an equilibrium distribution at temperature $T_i$, different from the temperature $T$ for the evolution at $t>0$. The distribution of the initial macroscopic density is the multidimensional version of~\eqref{eq:FinitSM}, and reads
\begin{equation}
    F[\rho(\boldsymbol{x},0)] \asymp \e^{-\Nsc^d F[\rho(\boldsymbol{x},0)]}
    \:,
    \quad
    F[\rho(\boldsymbol{x},0)]
    =
    \frac{1}{k_B T_i}
    \int \dd^d \boldsymbol{x} \left[
    f_i(\rho(\boldsymbol{x},0)) - f_i(\bar\rho)
- (\rho(\boldsymbol{x},0) - \bar\rho) f_i'(\bar\rho)
    \right]
    \:,
\end{equation}
where $f_i(\rho)$ is the free energy density for an infinite system at equilibrium at temperature $T_i$ and density $\rho$. The generating function of the correlations is
\begin{equation}
    \ln  \moy{\e^{\Nsc^d \int \lambda(\boldsymbol{x}) \rho(\boldsymbol{x},0) \dd^d \boldsymbol{x}}} \underset{\Nsc \to \infty}{\simeq}
    \Nsc^{d} G[\lambda]
    \:,
\end{equation}
where
\begin{equation}
    G[\lambda] = \int \lambda(\boldsymbol{x}) \rho^\star(\boldsymbol{x}) \dd^d \boldsymbol{x}
    - F[\rho^\star(\boldsymbol{x})]
    \:,
\end{equation}
with $\rho^\star$ determined by the saddle-point equation
\begin{equation}
    \label{eq:MinFSM}
    \frac{\delta F[\rho]}{\delta \rho(\boldsymbol{x})} \Bigg|_{\rho^\star} = \lambda(\boldsymbol{x})
    \quad \Rightarrow \quad
    \lambda(\boldsymbol{x}) =
    \frac{1}{k_B T_i} \left(
    f_i'(\rho^\star(\boldsymbol{x})) - f_i'(\bar\rho)
    \right)
    = \int_{\bar\rho}^{\rho^\star(\boldsymbol{x})}
    \frac{2 D_i(r)}{\sigma_i(r)} \dd r
    \:,
\end{equation}
where we have used the relation between the free energy and the transport coefficients~\eqref{eq:RelDSigmaFSM}. We thus have that
\begin{equation}
    \moy{ \rho(\boldsymbol{z},0) \rho(\boldsymbol{y},0)}_c
    \equiv \Nsc^{-d} \frac{\delta^2 G[\lambda]}{\delta \lambda(\boldsymbol{z}) \delta \lambda(\boldsymbol{x})}
    = \Nsc^{-d}\frac{\delta}{\delta \lambda(\boldsymbol{z})}
    \left[
        \rho^\star(\boldsymbol{x})
        - \int \dd^d \boldsymbol{y} \frac{\delta \rho^\star(\boldsymbol{y})}{\delta \lambda(\boldsymbol{x})}
        \left(
        \frac{\delta F[\rho]}{\delta \rho(\boldsymbol{y})} \Bigg|_{\rho^\star} - \lambda(\boldsymbol{y})
        \right)
    \right]
    \:.
\end{equation}
The second term vanishes due to~\eqref{eq:MinFSM}, thus
\begin{equation}
    \moy{ \rho_1(\boldsymbol{z},0) \rho_1(\boldsymbol{y},0)}_c
    =
    \Nsc^d \moy{ \rho(\boldsymbol{z},0) \rho(\boldsymbol{y},0)}_c
    =\frac{\delta \rho^\star(\boldsymbol{x})}{\delta \lambda(\boldsymbol{z})}
    = \frac{\sigma_i(\bar\rho)}{2 D_i(\bar\rho)} \delta(\boldsymbol{x} - \boldsymbol{z})
    \:,
\end{equation}
where we have again used~\eqref{eq:MinFSM}. Inserting this result into the expression~\eqref{eq:CorrelDensDens0SM}, we obtain
\begin{equation}
    \moy{\rho(\boldsymbol{x},t) \rho(\boldsymbol{y},0)}_c
    =
    \Nsc^{-d}\moy{\rho_1(\boldsymbol{x},t) \rho_1(\boldsymbol{y},0)}_c
    =
    \frac{\sigma_i(\bar\rho)}{2 D_i(\bar\rho)}
    \frac{\e^{- \frac{(\boldsymbol{x}-\boldsymbol{y})^2}{4 D(\bar\rho) t}}}{(4\pi D(\bar\rho) \Nsc^2 t)^{d/2}}
    \:.
\end{equation}
This expression holds for the macroscopic density $\rho(\boldsymbol{x},t)$. It implies that the microscopic density  $\rho_0(\boldsymbol{x},t) = \rho(\Nsc^{-1}\boldsymbol{x},\Nsc^{-2} t)$ satisfies at large times and large distances
\begin{equation}
    \moy{\rho_0(\boldsymbol{x},t) \rho_0(\boldsymbol{y},0)}_c
    \underset{\Lambda \to \infty}{\simeq}
    \frac{\sigma_i(\bar\rho)}{2 D_i(\bar\rho)}
    \frac{\e^{- \frac{(\boldsymbol{x}-\boldsymbol{y})^2}{4 D(\bar\rho) t}}}{(4\pi D(\bar\rho) t)^{d/2}}
    \:,
\end{equation}
which is Eq.~(16) of the main text.
Taking the Fourier transform of this expression yields the dynamical structure factor,
\begin{equation}
    \label{eq:StrFacSM}
    S(\boldsymbol{k},t) \equiv \int \frac{\dd^d \boldsymbol{x}}{\bar\rho} \: \e^{\I \boldsymbol{k} \cdot \boldsymbol{x}} \moy{\rho_0(\boldsymbol{x},t) \rho_0(\boldsymbol{0},0)}_c
    = \frac{\sigma_i(\bar\rho)}{2 \bar\rho D_i(\bar\rho)}
    \e^{- \boldsymbol{k}^2 D(\bar\rho) t}
    \:,
\end{equation}
which will be useful to obtain $D(\bar\rho)$ and $\sigma(\bar\rho)$ in the numerical simulations (see below).

\section{Numerical simulations}

\subsection{One dimensional systems}

The simulations in one dimension are performed by directly solving~\eqref{eq:EqBrownianPartSM}, discretised in time with a step $\delta t$, with $N$ particles on a ring of length $L = N/\rho$.
To initialise the system in an equilibrium configuration at temperature $T_i$, we start from a given initial configuration, for instance equally spaced particles (corresponding to $T_i = 0$ for the repulsive potentials considered here), and then perform $M$ steps of a Markov chain Monte-Carlo algorithm. In each of these steps, one particle is selected at random, and displaced by a distance $\varepsilon$. The energy change $\Delta E$ of this displacement is computed, and the move is accepted with probability $\min(1 , \e^{- \Delta E/(k_B T_i)})$. We typically perform $M=10^9$ such steps for a system with $N=400$ particles. This is faster than performing a time evolution with~\eqref{eq:EqBrownianPartSM} until the system has relaxed to equilibrium.

After this thermalisation step, the time evolution with~\eqref{eq:EqBrownianPartSM} is performed, keeping track of the displacement of each particle. At a given time $t$, the density of the system is constructed in the reference frame of each particle, and the observables $X_t$, $X_t \rho_0(X_t+x,t)$, etc... are computed. Their values are averaged over all the particles, as each particle can be treated as a tracer.

\begin{figure}
    \centering
    \includegraphics[width=0.35\textwidth]{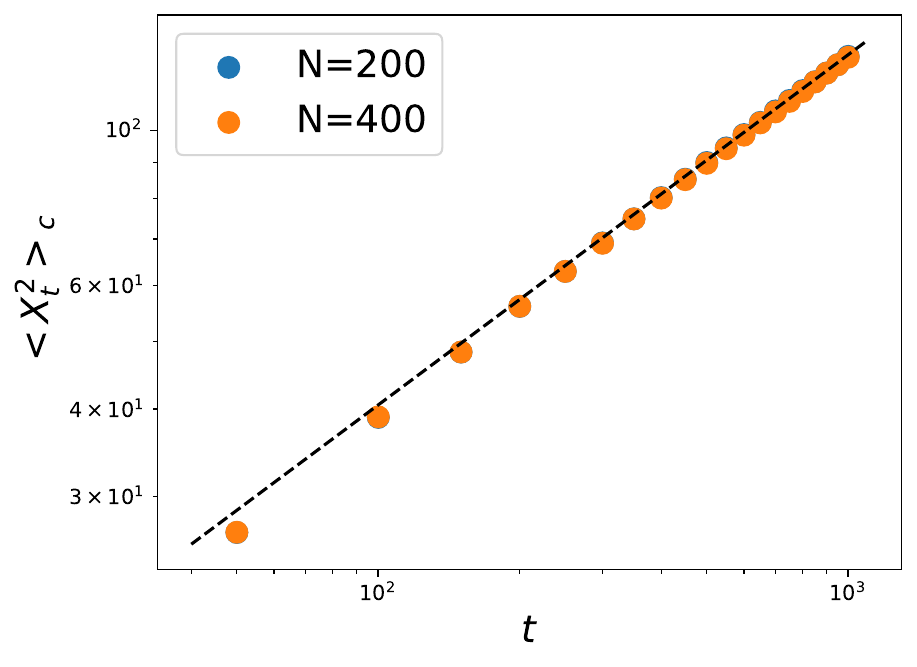}
    \includegraphics[width=0.35\textwidth]{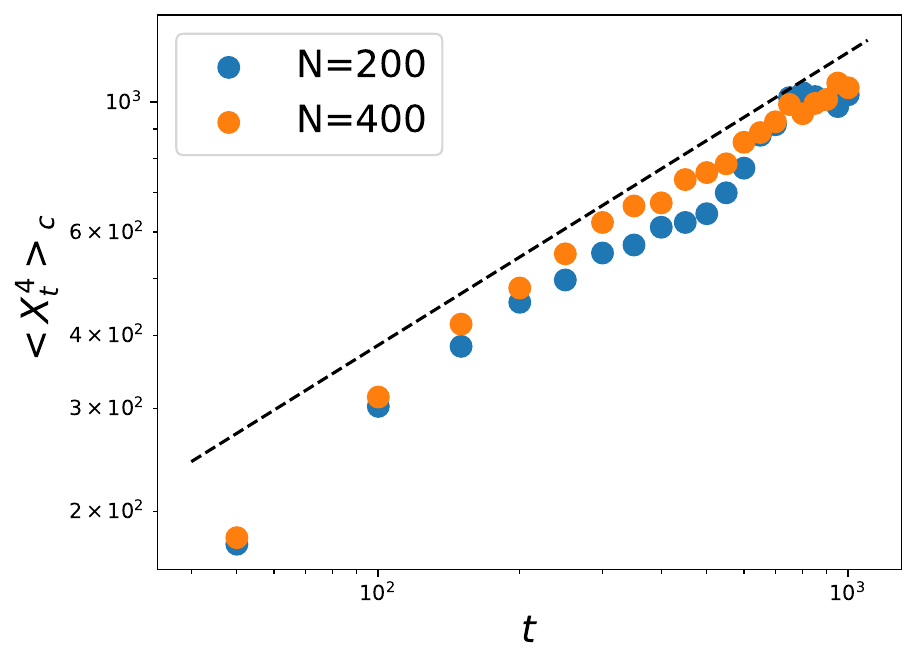}
    \caption{Numerical computation of the second cumulant $\moy{X_t^2}_c$ (left) and fourth cumulant $\moy{X_t^4}_c$ (right) for the Calogero potential as a function of time for different number of particles in the system. The parameters are $\rho = 0.2$, $T=1$, $\mu_0 = 1$, $k_B = 1$ and $a=1$. The results are averaged over $10^4$ simulations.}
    \label{fig:ConvCumulCalo}
\end{figure}

The numerical computation of the cumulants $\moy{X_t^n}_c$ becomes more difficult as the order $n$ is increased, for several reasons. (i) The fourth cumulant $\moy{X_t^4}_c$, which behaves as $\sqrt{t}$, is the difference of two terms that grow linearly with time, $\moy{X_t^4}_c = \moy{X_t^4} - \moy{X_t^2}^2$. This requires a high precision on each term, and thus more averaging.
(ii) Looking at Figs.~1 and~2 of the main text, $\moy{X_t^4}_c$ seems to reach its asymptotic behavior at a time much longer than the one
required by
$\moy{X_t^2}$. This requires running longer simulations.
(iii) The fourth cumulant $\moy{X_t^4}_c$ is also more sensitive to finite-size effects than the second cumulant $\moy{X_t^2}_c$, as illustrated in Fig.~\ref{fig:ConvCumulCalo}.
All these points result in a larger noise on the fourth cumulant, with
finite-size effects
which persist
even for $N=400$ particles. This is the origin of the deviation of the numerical results from the analytical predictions in the insets of Figs.~1 and~2 of the main text.

\bigskip

\begin{figure}
    \centering
    \includegraphics[width=0.4\textwidth]{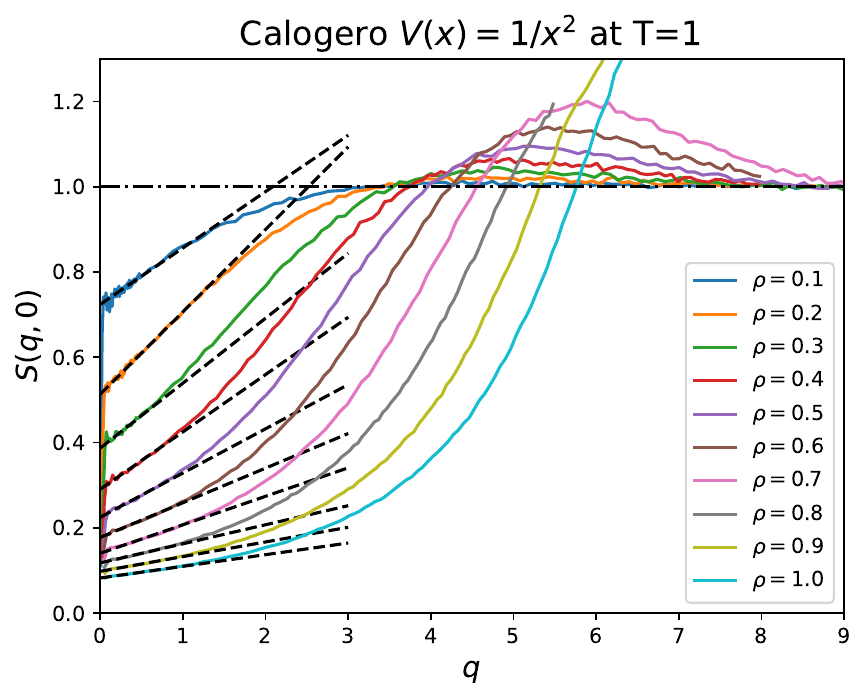}
    \includegraphics[width=0.4\textwidth]{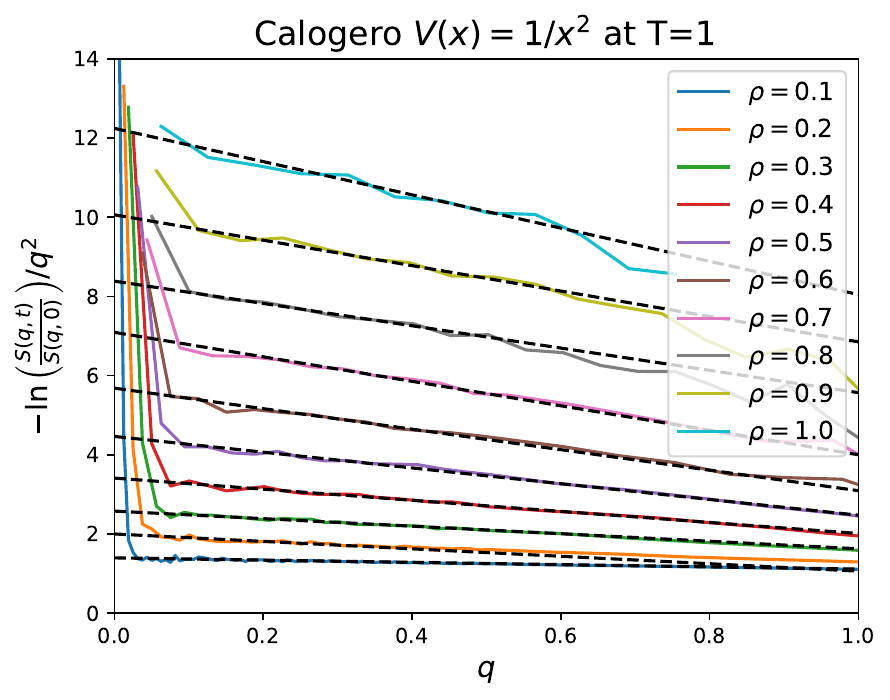}
    \caption{Numerical computation of the static structure factor $S(q,0)$ (left), and the exponential decay of the dynamical structure factor $S(q,t)$ at $t=1$ (right) for the Calogero potential~\eqref{eq:CalogeroSM} with $a=1$. The dashed lines are linear fits, from which the value at $q=0$ is obtained. The simulations are performed with $N = 100$ particles on a periodic ring of length $L = N/\rho$, up to time $t=100$ at temperature $T=1$. Since these are equilibrium quantities, we averaged them over the simulation time.}
    \label{fig:StrFacCalo}
\end{figure}

\begin{figure}
    \centering
    \includegraphics[width=0.4\textwidth]{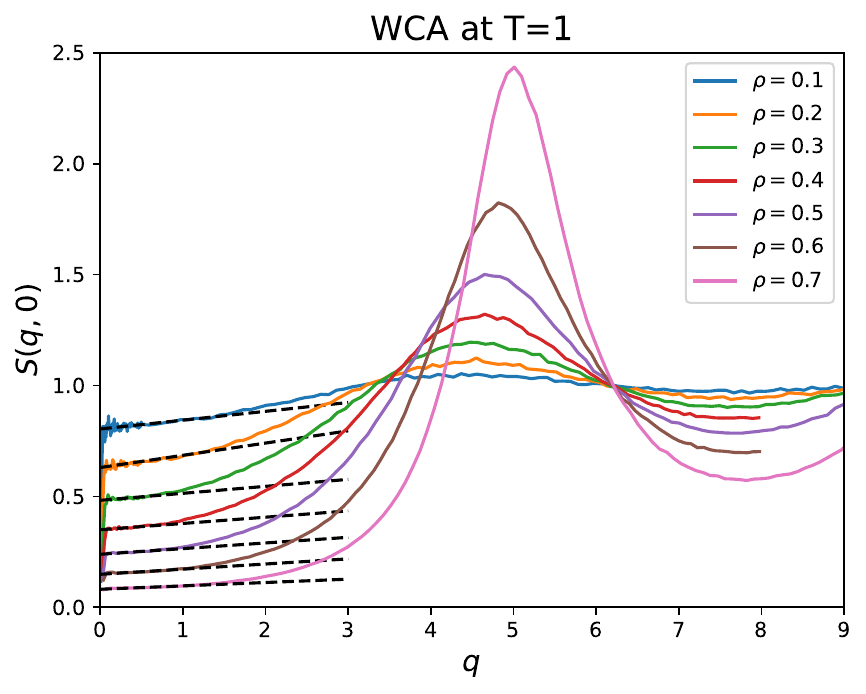}
    \includegraphics[width=0.4\textwidth]{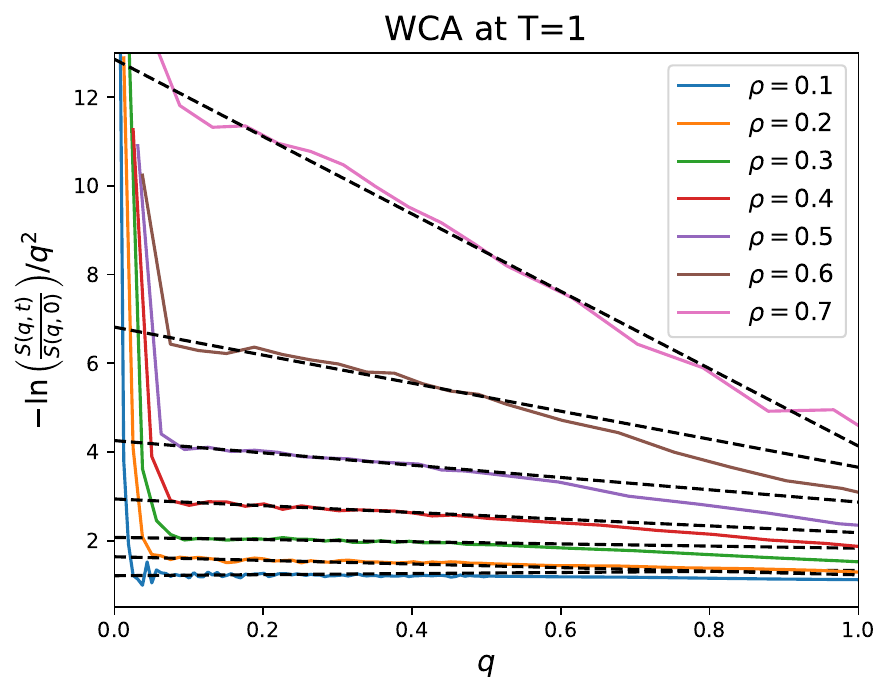}
    \caption{Numerical computation of the static structure factor $S(q,0)$ (left), and the exponential decay of the dynamical structure factor $S(q,t)$ at $t=1$ (right) for the WCA potential~\eqref{eq:WCASM} with $\ell=1$ and $a=1$. The dashed lines are linear fits, from which the value at $q=0$ is obtained. The simulations are performed with $N = 100$ particles on a periodic ring of length $L = N/\rho$, up to time $t=100$ at temperature $T=1$. Since these are equilibrium quantities, we averaged them over the simulation time.}
    \label{fig:StrFacWCA}
\end{figure}

\begin{figure}
    \centering
    \includegraphics[width=0.4\textwidth]{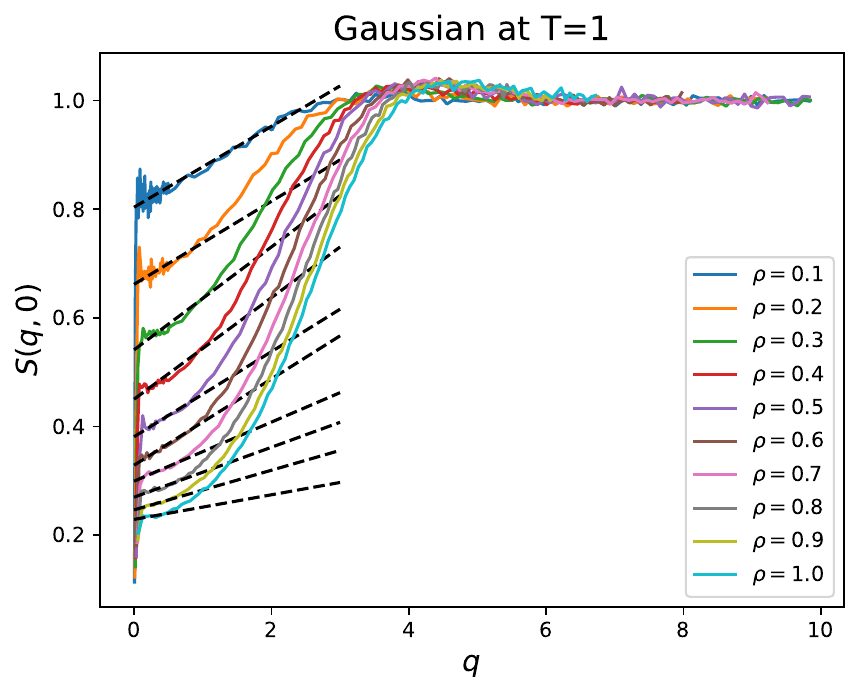}
    \includegraphics[width=0.4\textwidth]{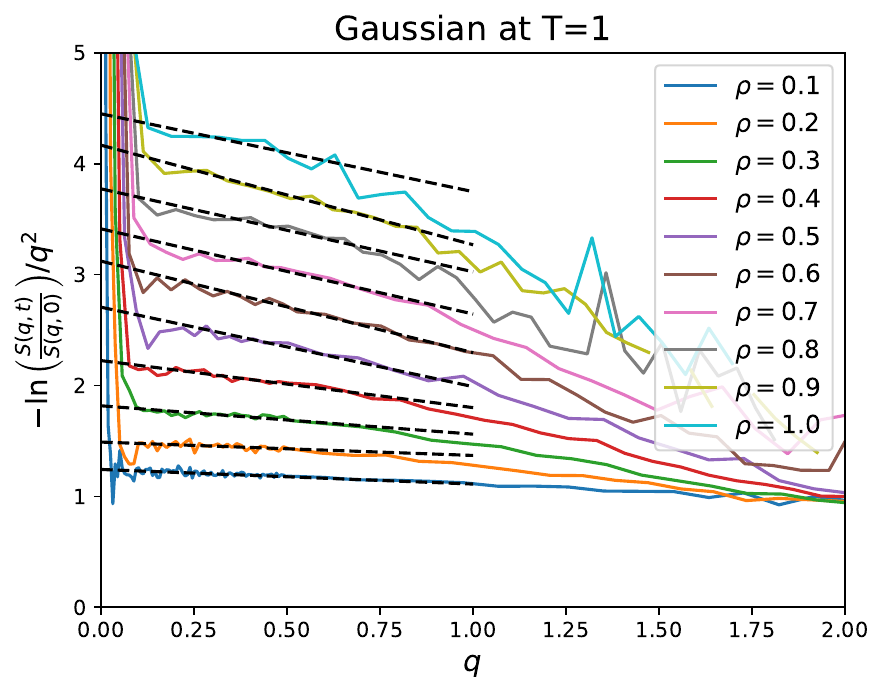}
    \caption{Numerical computation of the static structure factor $S(q,0)$ (left), and the exponential decay of the dynamical structure factor $S(q,t)$ at $t=1$ (right) for the Gaussian potential~\eqref{eq:GaussPotSM} with $a = 2$ and $\ell = 1$. The dashed lines are linear fits, from which the value at $q=0$ is obtained. The simulations are performed with $N = 100$ particles on a periodic ring of length $L = N/\rho$, up to time $t=100$ at temperature $T=1$. Since these are equilibrium quantities, we averaged them over the simulation time.}
    \label{fig:StrFacGaussian}
\end{figure}

\begin{figure}
    \centering
    \includegraphics[width=0.32\textwidth]{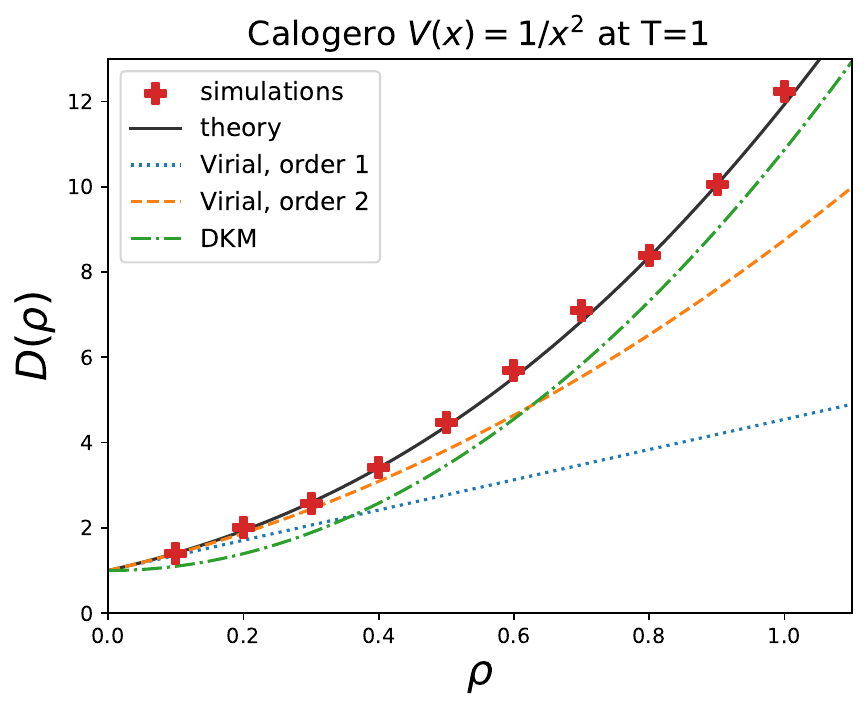}
    \includegraphics[width=0.32\textwidth]{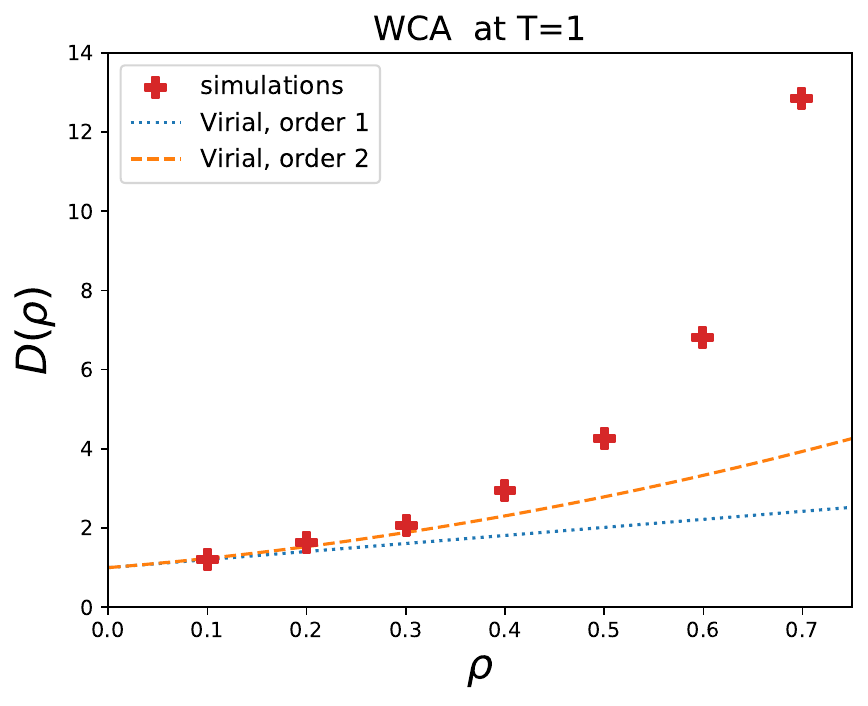}
    \includegraphics[width=0.32\textwidth]{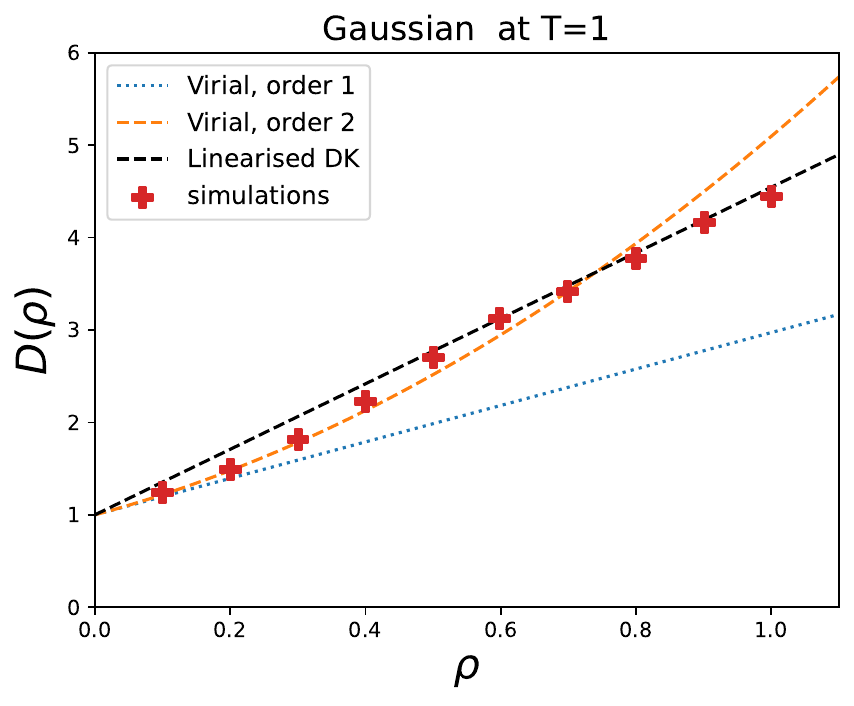}
    \caption{Diffusion coefficient $D(\rho)$ in dimension $d=1$ at $T=1$, for $\mu_0 = 1$ and for the different interaction potentials considered here. On the left plot, for the Calogero potential, the green dashed line is the prediction found in Ref.~\cite{Dandekar:2023SM} (labelled ``DKM''), obtained from the Dean-Kawasaki equation, which clearly deviates from the numerical simulations, while our prediction is in perfect agreement.}
    \label{fig:DiffCoefsSim}
\end{figure}

\begin{figure}
    \centering
    \includegraphics[width=0.4\textwidth]{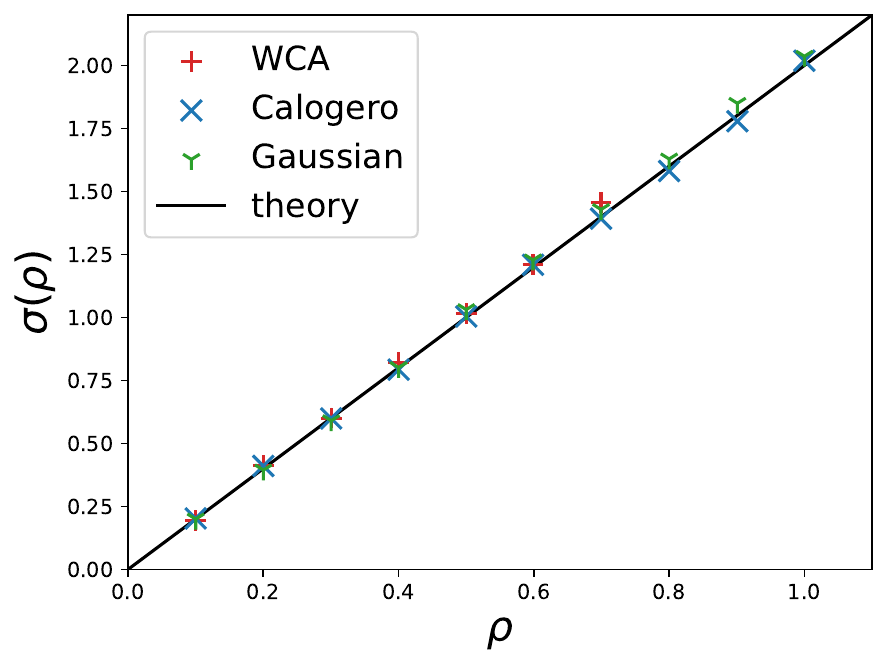}
    \caption{Mobility coefficient $\sigma(\rho)$ in dimension $d=1$  at $T=1$ and $\mu_0 = 1$, for the different interaction potentials considered here.}
    \label{fig:SigmaSim}
\end{figure}

To validate the expressions of $D(\rho)$~\eqref{eq:DSM} and $\sigma(\rho)$~\eqref{eq:SigmaSM}, as well as to estimate the range of validity of the virial expansion~\eqref{eq:PressureVirialSM}, we compute the values of $D(\rho)$ and $\sigma(\rho)$ in the simulations using the method of~\cite{Kollmann:2003SM}. We compute the dynamical structure factor
\begin{equation}
    \label{eq:DefStMicroSM}
    S(k,t) \equiv \frac{1}{N}
    \moy{ \sum_{i, j = 1}^N \e^{\I k (x_i(t) - x_j(0))}}
    \:,
\end{equation}
which behaves for small $k$ as~\cite{Kollmann:2003SM}
\begin{equation}
    \label{eq:StSM}
    S(k,t) \underset{k \to 0}{\simeq} S(\bar\rho) \e^{- k^2 D(\bar\rho) t}
    \:.
\end{equation}
Identifying with~\eqref{eq:StrFacSM} for the annealed case $T_i = T$, we get that
\begin{equation}
    \label{eq:S0SM}
    S(\bar\rho) = \frac{\sigma(\bar\rho)}{2 \bar\rho D(\bar\rho)}
    \:,
\end{equation}
as found in~\cite{Krapivsky:2015aSM}. Equations.~(\ref{eq:StSM},\ref{eq:S0SM}) give a way to compute $D(\rho)$ and $\sigma(\rho)$ in the simulations, from the evaluation of~\eqref{eq:DefStMicroSM}. From the behaviour of $S(k,0)$ for $k \to 0$, we obtain $S(\bar\rho)$, and thus the ratio $D(\bar\rho)/\sigma(\bar\rho)$~\eqref{eq:S0SM}. The diffusion coefficient is obtained from~\eqref{eq:StSM}, written as
\begin{equation}
    -\frac{1}{k^2} \ln \left(\frac{S(k,t)}{S(k,0)} \right)
    \underset{k \to 0}{\simeq} D(\bar\rho) t
    \:.
\end{equation}
The same method was used in Ref.~\cite{Poncet:2021SM}. This procedure is illustrated in Fig.~\ref{fig:StrFacCalo} for the Calogero potential~\eqref{eq:CalogeroSM}, in Fig.~\ref{fig:StrFacWCA} for the WCA potential~\eqref{eq:WCASM}, and in Fig.~\ref{fig:StrFacGaussian} for the Gaussian potential.
The results for $D(\rho)$ are shown in Fig.~\ref{fig:DiffCoefsSim} and for $\sigma(\rho)$ in Fig.~\ref{fig:SigmaSim}.

\subsection{Two dimensional systems}
\label{sec:simul-2d}
\begin{figure}
    \centering
    \includegraphics[width=0.45\textwidth]{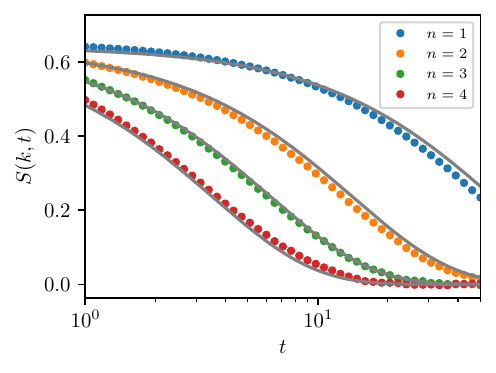}
    \put(-130,170){Gaussian, $\rho=0.2$}
    \includegraphics[width=0.45\textwidth]{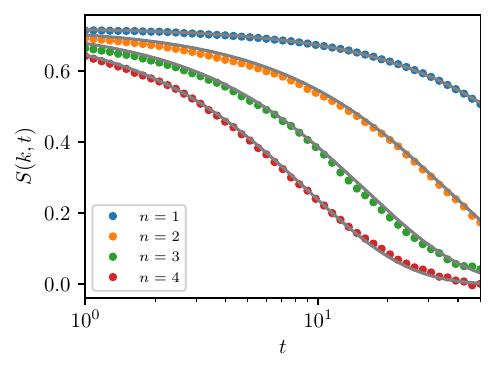}
    \put(-130,170){WCA, $\rho=0.1$}
    \caption{Dynamical structure factor $S(k,t)$ of two-dimensional systems of overdamped Brownian particles with either Gaussian (left) or WCA (right) interaction potentials, as measured in numerical simulations (see Sec.~\ref{sec:simul-2d} for the details). At time $t=0$, the temperature is suddenly quenched from $T_i=1$ to $T$ (with $T=0.1$ or $T=0.5$ in the Gaussian and WCA cases, respectively).
    Measured values for $k=2\pi n/L$ (symbols) are shown together with the corresponding analytical MFT prediction~\eqref{eq:StrFacSM} (grey lines).
    }
    \label{fig:2d-simul}
\end{figure}

Brownian dynamics simulations in $d=2$ were performed using the LAMMPS computational package \cite{Thompson2022SM}, and PyLammps, the wrapper Python class for LAMMPS.
To validate the prediction for the dynamical structure factor $S(\boldsymbol{k},t) $ in Eq.~\eqref{eq:StrFacSM}, we aim at measuring its particle-based discrete version given in Eq.~\eqref{eq:DefStMicroSM}.
The measurements presented in Fig.~\ref{fig:2d-simul} were obtained by choosing either Gaussian or WCA interaction potentials, see Eqs.~\eqref{eq:GaussPotSM} and~\eqref{eq:WCASM}, respectively. We simulated $N$ particles at density $\rho$, in a periodic box of linear size $L=\sqrt{N/\rho}$.

After letting the system thermalize
for a time $t_\text{term}$ (with time step $\Delta t=10^{-4}$) at the initial temperature $T_i=1$, we change the temperature to $T$ and measure $S(k,t)$ at logarithmically distributed time intervals. The system being identical along the $x$ and $y$ directions, we calculate $S(k,t)$ by letting $\boldsymbol{k}$ assume either of these two orientations, and then take their average. (Note that a perfect isotropy along the other directions is not realized in a finite system in a square box.) We measured $S(k,t)$ for $k=2\pi n/L$ with $n$ integer, as appropriate for a system in a periodic box.
We repeated this procedure for $n_\mathrm{tot}=10^4$ uncorrelated samples, and finally took their average.

For Gaussian interaction potentials, we used $N=200$, $\rho=0.2$, $T=0.1$ and $t_\mathrm{term}=200$, whereas for WCA potentials we used $N=400$, $\rho=0.1$, $T=0.5$ and $t_\mathrm{term}=1000$ (note that both finite-size effects and long-wavelength fluctuations are more pronounced in the latter case).

\bibliographystyle{apsrev4-1}

\end{document}